\documentclass{article}

\usepackage[utf8]{inputenc}
\usepackage[english]{babel}
\usepackage{eurosym}
\usepackage{xcolor}
\usepackage{mathrsfs}
\usepackage{graphicx}
\usepackage{float}
\usepackage{bbold}
\usepackage{bm}
\usepackage{cite}
\usepackage{amssymb,amsfonts, amsmath}
\usepackage{textcomp}
\usepackage[T1]{fontenc}
\usepackage{lmodern}
\usepackage{siunitx}
\usepackage{caption}
\usepackage{authblk}
\usepackage{subcaption}
\usepackage[colorlinks=true, allcolors=blue]{hyperref}
\usepackage[normalem]{ulem}
\usepackage[a4paper,top=2cm,bottom=2cm,left=3cm,right=3cm,marginparwidth=1.75cm]{geometry}

\newcommand{\zbar}{{\overline{z}}}
\newcommand{\Ybar}{{\overline{Y}}}
\newcommand{\Lbar}{{\overline{L}}}
\newcommand{\Sbar}{{\overline{S}}}
\newcommand{\etabar}{{\overline{\eta}}}
\newcommand{\lambdabar}{{\overline{\lambda}}}
\newcommand{\Kbar}{{\overline{K}}}

\newcommand{\vbar}{{\overline{v}}}
\newcommand{\ibar}{{\overline{i}}}

\title{Complex Couplings - A universal, adaptive and bilinear formulation of power grid dynamics}
\author[1]{Anna Büttner \footnote{buettner@pik-potsdam.de}}
\author[1]{Frank Hellmann \footnote{hellmann@pik-potsdam.de}}
\affil[1]{Potsdam-Institute for Climate Impact Research, 14473 Potsdam, Germany}
\date{}
\begin{document}
\maketitle

	The paper is now published in PRX Enegry \cite{PRXEnergy.3.013005}. Please refer to the PRX version from now on. \\

	Anna Büttner and Frank Hellmann. "Complex Couplings-A Universal, Adaptive, and Bilinear Formulation of Power Grid Dynamics." PRX Energy 3 (2023): 013005. \url{https://journals.aps.org/prxenergy/abstract/10.1103/PRXEnergy.3.013005}

\section*{Abstract} 
	
    The energy transition introduces new classes of dynamical actors into the power grid. There is especially a growing need for so-called grid-forming inverters (GFIs) that can contribute to dynamic grid stability as the share of synchronous generators decreases. Understanding the collective behavior and stability of future grids, featuring a heterogeneous mix of dynamics, remains an urgent and challenging task.
    
    Two recent advances in describing such modern power grid dynamics have made this problem more tractable: First, the normal form for grid-forming actors provides a uniform, technology-neutral description of plausible grid dynamics, including grid-forming inverters and synchronous machines. Secondly, the notion of the complex frequency has been introduced to effortlessly describe how the nodal dynamics influence the power flows in the grid.

    The major contribution of this paper is to show how the normal form approach and the complex frequency dynamics of power grids combine, and how they relate naturally to adaptive dynamical networks and control affine systems.
    
    Using normal form and complex frequency, we derive a remarkably elementary and universal equation for the collective grid dynamics. Notably, we obtain an elegant equation entirely in terms of a matrix of complex couplings, in which the network topology does not explicitly appear. These complex couplings give rise to new adaptive network formulations of future power grid dynamics. We give a new formulation of the Kuramoto model with inertia as a special case.

    Starting from this formulation of the grid dynamics, the question of the optimal design of future grid-forming actors becomes treatable by methods from affine and bilinear control theory. We demonstrate the power of this perspective by deriving a quasi-local control dynamics that can stabilize arbitrary power flows, even if the effective network Laplacian is not positive definite.

\section{Introduction} 
    Due to the energy transition, conventional power plants are replaced by renewable energy sources (RES) such as wind and solar PV. RES are predominantly connected to the power grid via power electronic inverters.
    Furthermore, power electronic interfaces are also in use for many loads, and for connecting storage solutions such as batteries to power grids. 
    Currently, most inverters are grid-following and have to rely on a stable grid to function. However, as we move towards a fully renewable grid, there is a growing need for so-called grid-forming inverters \cite{ensoe_grid_forming_2020} (GFIs) which can contribute to the grid stability independently of conventional generation. 
    
    Popella et al. have shown that following one of the scenarios in the German Network Development Plan \cite{bna_scenarios_2020}, more than 80$\%$ of all inverters installed from 2021 need to be grid forming to assure a stable power grid in 2035 \cite{popella_grid_forming_2021}. This shift presents a significant challenge, as GFIs are a relatively new and complex technology, of which there is a limited practical and theoretical understanding. In future grids, these GFI will coexist with a much reduced number of conventional generators. The collective dynamics of such inverter-dominated networks remain a challenging topic \cite{prevost2019deliverable}. 
    
    For almost all proposed types of GFIs, there is an analysis of the stability of the individual machine. The stability of the entire network, however, is rarely addressed, with \cite{schiffer_droop_2014} and \cite{colombino_dvoc_2017} as notable exceptions. Most research on the collective behavior of the power grid is still based on the paradigmatic Kuramoto model \cite{witthaut2022collective}. For generators, we already know from the study of higher-order models that include voltage dynamics, that the Kuramoto Model can be misleading, for the individual dynamics \cite{machowski_power_2008} but also for the collective phenomena \cite{auer2016impact}. As we move towards an inverter-dominated grid, we even lack the consensus on higher-order models of GFIs, as vastly different realizations of grid-forming capabilities are possible. However, important phenomena occurring in inverter-dominated networks can not be captured by the Kuramoto model.
    
    To study collective phenomena in heterogeneous systems, a description of all grid-forming actors, conventional and future GFI-based, is needed. Kogler et al. \cite{kogler_normalform_2022} address this, by introducing the normal form, a dynamic technology-neutral formulation for the dynamics of grid-forming actors. This normal form approach provides a concise and meaningful parametrization of the space of plausible power grid actors, especially when they are near their desired operational state. Thus, it is well suited to the study of power systems with a heterogeneous mix of dynamical actors.
    
    Independently, Milano recently introduced the notion of the complex frequency \cite{milano_complex_frequency_2022} to elegantly describe how the nodal dynamics influence the power flows in the grid.

    In this paper, we show that the two formulations introduced above complement each other perfectly. Together, they provide an elementary and universal formulation of the space of plausible power grid dynamics. Further, by focusing on the evolution of the couplings between nodes, we can derive a new and surprisingly elegant formulation of the collective grid dynamics. This formulation has the remarkable property that the network adjacency matrix no longer explicitly appears in the equations, but is encoded in the initial condition.
    
    This immediately gives rise to an adaptive network formulation of the power grid dynamics that is different from that employed in \cite{berner_2021_adaptive_powergrids}. As a special case, we derive the Kuramoto model with inertia in this formulation, but the general formulations provided again make no simplifying assumptions such as lossless coupling or specific voltage dynamics.

    The power and voltage magnitude square, the most important control objectives of power grids, are linear combinations of the complex coupling. Thus, if we use the complex coupling dynamics as the dynamics to be controlled, we obtain an elegant quadratic-bilinear tracking problem. Solving this problem under the constraint of a fully decentral controller using only local information is challenging; however, the formulation provides a natural quasi-local Lyapunov-Function-Controller that is globally synchronizing while reducing the tracking error.

    A major contribution of this paper is to systematically collect and describe the various ways in which complex frequency dynamics of power grids, normal form descriptions of grid-forming actors, and concepts from control theory and dynamical systems intersect. To provide the reader with a background of the relevant fields we will begin with recalling the most relevant concepts from power grids, adaptive dynamical networks, and bilinear control systems in section \ref{sec:background}. In the second part \ref{sec:bilinear_coupling} we introduce the new formulations of the grid dynamics. Finally, we explore the control aspects in section \ref{sec:control_problems}. We provide an appendix with various useful related calculations.
    
    \section{Theoretical Background}
    \label{sec:background}
    \subsection{Balanced AC Power Grid Variables}
    Generally, power grids consist of several parallel circuits carrying phase-shifted AC voltages and currents. We will focus on the typical case of three phases. A three-phase signal $V_1$, $V_2$, $V_3$ is called balanced if it satisfies $V_1 + V_2 + V_3 = 0$ at all times. Balanced signals can be expressed in terms of a rotating 2D vector, often referred to as $dq$-coordinates \cite{Power_System_Dynamics}. Defining the complex state $v = v_d + j v_q$, also called the Park vector, we can write the original signal as $V_l = \mathrm{Re}(\exp{(j l 2/3 \pi ) v})$ where $l$ depicts the different phases. The complex voltage $v$ can be written in terms of phase $\theta$ and amplitude $\mathbf{V} \geq 0$ as:
    \begin{align}
        v(t) &= \mathbf{V} e^{j \theta} \label{eq:voltage_phasor}
    \end{align}
    Note that, with our definition, the amplitude $\mathbf{V}$ is the (peak) voltage magnitude of the three-phase signal. As we will assume balanced conditions throughout, the Park vector $v(t)$ captures the full state space of the system. The same procedure is used to express the three-phase currents $I_h$ in terms of a single complex current $i(t)$.

    It is then a straightforward calculation to see that the real part of $v \overline i$ is $\frac32$ times the instantaneous power contained in a three-phase current $i$ at voltage $v$. 
    The imaginary part is called the instantaneous reactive power and is interpreted as the amount of energy that is injected into the transmission line without flowing through it. 
    The full quantity $S_c = \frac23 v \overline i$ is called the complex power. In what follows we will absorb the factor $\frac23$ into the coefficients and abuse terminology slightly to speak simply of 
    \begin{align}
    S = P + j Q = v \overline i
    \end{align}
    as the complex power.
    
\subsection{Complex Frequency}
    \label{sec:nf}
    The frequency $f$ is the most important variable for operating, controlling, and monitoring power grids \cite{Power_System_Dynamics}. According to the accepted IEEE standard \cite{ieee_frequency_def_2018} the frequency $f(t)$ of a signal $x(t)$ with amplitude $\mathbf{X}$ and phase $\phi$ is defined as:
    \begin{align}
        x(t) &= \mathbf{X}(t) \cos(\phi(t)) \\
        f(t) &= \frac{1}{2\pi} \dot{\phi}(t).
    \end{align}
    This standard definition of frequency in AC power grids is only meaningful when a state with slowly varying amplitude is assumed. However, in power grids, fast amplitude changes in the voltage are possible. This standard definition can not distinguish between phase and amplitude variations in a meaningful way. The objective of the complex frequency, which has been introduced in \cite{milano_complex_frequency_2022}, is to overcome this issue by providing an interpretation of the instantaneous frequency as a state space velocity. As long as it is non-zero, we can consider the time derivative of the logarithm of the voltage $v(t)$ \eqref{eq:voltage_phasor}:
    \begin{align}
        v(t) &= e^{\operatorname{ln}(\mathbf{V}) + j \theta} = e^{\sigma + j \theta} \\
        \eta &:= \dot{\sigma} + j \dot{\theta} = \rho + j \omega \label{eq:complex_frequency} \\
        \dot{v}(t) &= (\dot{\sigma} + j \dot{\theta}) \cdot v = (\rho + j \omega) \cdot v = \eta \cdot v \label{eq:voltage_dynamics}
    \end{align}
    here $\omega$ is the angular velocity of the signal and $\rho$ is the rate of change of the log-voltage amplitude. The complex number $\eta$ is defined as the complex frequency \cite{milano_complex_frequency_2022}. For slow-changing amplitudes, where both definitions are applicable, the standard definition of the frequency coincides with the imaginary part of the complex frequency.
    
    The complex frequency has already found numerous applications ranging from the inertia estimation of virtual power plants \cite{zhong_interia_2022} to analyzing the linear \cite{he_complex_frequency_sync_2022} and nonlinear stability \cite{he_non_linear_stability_2023} of networks which are composed of dispatchable virtual oscillators \cite{seo_dvoc_2019}, a proposed type of GFI. Furthermore, \cite{moutevelis_taxonomy_2022} shows the connection between the complex frequency and power for several different converter controllers, including GFIs.
    
    The authors of \cite{he_complex_frequency_sync_2022} introduce complex frequency synchronization, a novel concept of synchronization. Equation \eqref{eq:complex_frequency} shows that to obtain full synchrony of the voltages, both the angular frequency and the rate-of-change-of-voltage have to match. During rate-of-change-of-voltage synchronization, the voltage amplitudes are still allowed to change, but at the same exponential rate. 
    \subsection{Normal Form}
    The authors of \cite{kogler_normalform_2022} have introduced the normal form, a technology-neutral formulation for the dynamics of grid-forming actors. The main result of \cite{kogler_normalform_2022} is that the symmetry under global phase shifts of the nodal dynamics can be exploited to formulate the dynamics in terms of invariants of the symmetry. The entire interaction between grid-forming actors and the grid can be written in terms of the active $P$ and reactive power $Q$, as well as the square of the voltage magnitude $\nu = v \vbar$, which form an elegant complete set of invariants for power grids. $P$, $Q$, and $\nu$ are exactly the quantities needed to describe the desired behavior of grid-forming actors, as these are responsible for balancing active power and providing enough reactive power to maintain the nominal voltage levels. Following \cite{kogler_normalform_2022} the dynamical behavior of grid-forming actors can be generally written in this form:
    \begin{align}
        \dot{v} &= v g^v(\nu, P, Q, x_c)\\        
        \dot{x}_c &= g^{x_{c}}(\nu, P, Q, x_c) \label{eq:internal_dynamics}
    \end{align}
    where $v$ is again the complex voltage, $x_c$ are the internal dynamics of the control of the grid forming actors and $g^v$, and $g^{x_c}$ are differentiable, potentially nonlinear, functions. Using the complex frequency \eqref{eq:voltage_dynamics} we see that we can rewrite the first equation simply as:
    \begin{align}
        \eta = g^v(\nu, P, Q, x_c) \label{eq:eta_normal_form}
    \end{align}
    which gives the dynamics a novel interpretation. The complex frequency $\eta = \frac{\dot v}{v}$ determines, by definition, the voltage dynamics. As $\eta$ is invariant under global phase shifts, it can only depend on other invariants if the entire system is to be invariant. In particular, $\eta$ can not depend explicitly on $v$.
    
    \subsection{Adaptive Dynamical Networks}
    \label{sec:adaptive_networks}
    Adaptive dynamical networks are a class of systems that change their coupling $K_{hm}(t)$ over time depending on the dynamical state $y_h$ of the nodes of the network. The review paper \cite{berner2023adaptive} gives an excellent overview of the applications, dynamical phenomena, and available mathematical methods for adaptive dynamical networks. Furthermore, \cite{Sawicki_adaptivity_2023} features a rundown of the concept of adaptivity and how it applies across different scientific disciplines. In this work, we will focus on weighted adaptive dynamical networks that are in general defined as:
    \begin{align}
        \dot{y}_h &= f_h(y_h) + \sum_{m=1}^{M} K_{hm} \gamma (y_h, y_m) \label{eq:dynamical_network} \\
        \dot{K}_{hm} &= H(K_{hm},y_h, y_m) \label{eq:coupling_adaptation}
    \end{align}
    where $f_h$ describes the local dynamics of node $h$, $\gamma$ is the coupling function and $H$ the is the adaptation function.
    
    Recently, it has been revealed that dynamical power grid models, based on phase oscillator models are connected to adaptive networks \cite{berner_2021_adaptive_powergrids}. In particular, it has been shown that the Kuramoto model with inertia and the Kuramoto model with inertia and voltage dynamics \cite{Schmietendorf_2014_third_order}, in electrical engineering referred to as a third-order machine model \cite{machowski_power_2008}, can be written as a system of adaptively coupled phase oscillators. Using this approach the voltage dynamics can be interpreted as an additional adaptivity term. It has been shown that phase oscillator models and oscillators on adaptive networks share many different dynamical phenomena such as solitary states \cite{berner_2021_adaptive_powergrids, hellmann_2020_lossy_coupling}.
    
    In this work, we will introduce a type of complex coupling $K_{hm}$ that allows us to represent general dynamical power grid models as adaptive networks as long as the complex frequency $\eta$ is well defined.
    
    \subsection{Bilinear and Control Affine Systems}
    \label{sec:bilinear}
    Power grids are inherently non-linear systems. Therefore, to control them, non-linear control theory with all its associated difficulties is required. Instead, a linear approximation of the system could be used, but then non-linear effects cannot be taken into account and the controller will have limited effectiveness.
    
    The complex frequency and normal form equations reveal that the power grid can be interpreted as a system in which the control variable is linear. Such systems are called control affine systems \cite{sontag_mathematical_control_theory_1998}, and are of the form:
    \begin{align}
        \mathbf{\dot{y}} = \mathbf{F(y)} + \mathbf{G(y) u} \label{eq:control_affine}
    \end{align}
    where $\mathbf{y}$ and $\mathbf{u}$ are the state and control vector, respectively.

    The power grid equations we will find are bilinear systems \cite{pardalos_control_bilinear_systems_2008}, meaning that $\mathbf{F(y)}$ and $\mathbf{G(y)}$ are linear functions of $y$. This leads to the bilinear form of the control system:
    \begin{align}
        \dot{y}(t) = \sum_{a=1}^A u_a(t) N_a \cdot y(t) \label{eq:bilinear_system}
    \end{align}
    where $y$ and $u$ are again the state and control vector, respectively and $N_a$ are the system matrices.

    Bilinear systems are an area of intense research as they occur in various contexts \cite{mohler_overview_1980}, such as nuclear reactors. There is considerable work on bilinear control systems \cite{pardalos_control_bilinear_systems_2008}. Below, we will see that exploiting the control affine structure already allows us to derive interesting new control laws. 
    
    \section{New Power Grid Equations}

    We begin by introducing the dynamics of the complex coupling, before deriving new adaptive formulations of power grids.
    
    \subsection{Complex Coupling Dynamics}
    \label{sec:bilinear_coupling}

    In \cite{milano_complex_frequency_2022} Milano considers the dynamics of the complex power $S_h$ in terms of a dynamical coupling term. In Milano's paper, these coupling terms are referred to as $s_{hm}$. To avoid confusion with the power flows on the lines, we will use $K_{hm}$ in this work. Take $Y_{hm} = \frac{1}{Z_{hm}}$ as the admittance on the line connecting nodes $h$ and $m$. We then introduce the admittance weighted graph Laplacian $L_{hm}$ as:
    \begin{align}
        L_{hm} &= - Y_{hm} + \delta_{hm} \sum_k Y_{hk}  
    \end{align}
     The dynamical coupling terms $K_{hm}$, that will play the role of the adaptive coupling in the sense of equation \eqref{eq:coupling_adaptation}, are defined as:
    \begin{align}
        K_{hm} &= v_h \Lbar_{hm} \vbar_m.
    \end{align} 
    where $\Lbar_{hm}$ is the complex conjugate of the Laplacian. These coupling terms $K_{hm}$ do not have a physical interpretation, but they have the important property that:
    \begin{align}
        S_h = \sum_m K_{hm} \label{eq:S_h_relation}   
    \end{align}
    holds, as:
    \begin{align}
        i_{hm} &= Y_{hm} (v_h - v_m) \label{eq:line-current}\\
        i_h &= \sum_m i_{hm} = \sum_m L_{hm} v_m\\
        S_{h} &= v_h \ibar_h = \sum_m v_h \Lbar_{hm} \vbar_m = \sum_m K_{hm}. \label{eq:power_coupling}
    \end{align}   
    Using the voltage dynamics \eqref{eq:voltage_dynamics} the dynamics of the coupling are given by:
    \begin{align}
        \dot K_{hm} = (\eta_h + \etabar_m) K_{hm}. \label{eq:coupling_dynamics}
    \end{align}
    Thus the dynamics for the nodal complex power $S_{h}$ is given by:
    \begin{align}
        \dot S_h = \eta_h S_h + \sum_m K_{hm} \etabar_m \label{eq:power_flow_dynamics_coupling}
    \end{align}
    which is equation (32) in \cite{milano_complex_frequency_2022}. The diagonal coupling terms $K_{mm}$ describe the evolution of the local voltage magnitude square $\nu_m$ up to a constant factor: 
    \begin{align}
        \nu_m &= v_m \vbar_m = \frac{K_{mm}}{\Lbar_{mm}} = \frac{K_{mm}}{\sum_h \Ybar_{mh}} \label{eq:nu_coupling}
    \end{align}
    From equation \eqref{eq:coupling_dynamics} we can see that the system has two possible sets of fixed points. The first set of fixed points coincides with $K_{hm} = 0$ which can either be achieved when $h$ and $m$ are not connected by a transmission line or when there is a total voltage collapse at either node. The second type of fixed point is defined by $\eta_h = - \etabar_m$ throughout the connected component. If the system is connected at the fixed point, that is, for every two nodes there is a walk of non-zero $K_{hm}$ connecting them, then this implies that $\eta_h = j \omega_{global}$ for all $h$.
    
    This shows an immediate advantage of considering a dynamical system given entirely in terms of invariants, the desirable operating states are fixed points rather than limit cycles, which simplifies system identification and control synthesis considerably. The stability of the synchronous fixed points can now be expressed using a master stability function \cite{pecora_master_1998}, an approach that has recently been extended to a large class of adaptive networks \cite{berner_desynchronization_2021}. 

    
    Following \cite{milano_complex_frequency_2022} we assumed that the transmission line dynamics are faster than the nodal dynamics and that a quasi-steady state model can be employed for the transmission lines. In Appendix~\ref{sec:line-dynamics} we show briefly how to extend this description to more realistic line models. Furthermore, we point out that a similar, albeit less concise formulation can be achieved in power flow variables. The derivation can be found in the appendix as well \ref{sec:power_flow_vars}. 

    The coupling dynamics \eqref{eq:coupling_dynamics} can be used to rewrite general power grid dynamics as adaptive dynamical networks, as introduced in section \ref{sec:adaptive_networks}, as long as the complex frequency dynamics are well-defined. In the following, we will perform this reformulation for the classical Kuramoto model with inertia \ref{sec:kuramoto_adaptive}. Using the normal form \ref{sec:coupling_NF}, rather than the dynamics of the complex frequency, we derive an elegant equation entirely in terms of the matrix of complex couplings, in which the network topology does not explicitly appear.
    
    These reformulations generally come at the price of a larger phase space, which is typical for adaptive networks. In general, the phase space has $4N + 4 E$ dimensions, where $N$ is the number of nodes and $E$ is the number of edges in the power grid. The normal form is a notable exception to this rule, as demonstrated in section \ref{sec:coupling_NF}, the phase space has a dimension of only $4 E$. 
    
    \subsection{Adaptive Kuramoto Model with Inertia}
    \label{sec:kuramoto_adaptive}
    The Kuramoto model with inertia is the classical model to describe a generator in a power grid \cite{machowski_power_2008}. It is used to describe the short-term behavior of the generator, also called the first swing. For this reason, it is also called the Swing equation in this context. It is given by:
    \begin{align}
        \dot{\theta}_h &= \omega_h \\
        \dot{\omega}_h &= \frac{1}{2H_h}(P_{h}^s - D_h \omega_h - \sum_m P_{hm}) \label{eq:swing_eq}
    \end{align}
    where $\theta_h$ and $P_{h}^s$ are the phase angle and active power set point of node $h$. $D_h$ is the damping coefficient and $H_h$ is the inertia constant.
    
    As the Kuramoto model with inertia has no voltage dynamics the complex frequency is purely imaginary $\eta_h = j \omega_h$. Using the relation between the coupling and the complex power \eqref{eq:S_h_relation} and the coupling dynamics \eqref{eq:coupling_dynamics} we can rewrite the Kuramoto model with inertia as the following adaptive network:
    \begin{align}
        \dot{\eta}_h &= \frac{1}{2 H_h} (jP_{h}^s - D_h \eta_h - \frac{j}{2} \sum_m (K_{hm} + \Kbar_{hm}))  \\
        \dot K_{hm} &= (\eta_h + \etabar_m) K_{hm}.
    \end{align}
    It is important to note that this form of adaptive power grid dynamics is not the same as the formulation introduced in \cite{berner_2021_adaptive_powergrids}. As the complex coupling dynamics given above are formulated entirely in terms of invariants of the limit cycle, the phases do not appear explicitly. Thus synchronization at arbitrary frequencies is represented here by convergence to different fixed points, rather than limit cycles. For cluster synchronization, where several frequencies coexist, the clusters are internally approximately at fixed points, while the coupling between them oscillates rapidly. This formulation thus automatically realizes a fast-slow separation of variables for these common asymptotic states. 
    \subsection{Self-contained complex coupling dynamics}
    \label{sec:coupling_NF}
    By considering the equations of $\dot \eta$, the normal form can be rewritten as an adaptive equation in a similar way to the Kuramoto model above. Instead, we can also directly combine the normal form equations for $\eta$ and the complex couplings $K_{km}$. These equations provide a self-contained dynamical system expressed entirely in terms of invariants of the limit cycle. For notational simplicity, we first neglect the internal variables. Then the entire power grid's dynamics are given by:
    \begin{align}
        \dot K_{hm} &= (\eta_h + \etabar_m) K_{hm} \nonumber \\
        \eta_h &= g_h(P_h, Q_h, \nu_h) \nonumber \\
        P_h + j Q_h &= \sum_m K_{hm} \nonumber \\
        \nu_m &= \frac{K_{mm}}{\Lbar_{mm}}.\label{eq:coupling_dynamics_system}
    \end{align}
    Remarkably this system of differential equations depends only on the diagonal of the Laplacian $L_{hm}$, i.e. the weighted degrees. The line admittances are only encoded in the initial conditions.
    
    If there are internal degrees of freedom of the normal form, then there are dynamical variables at the node as well. Further, expressing $g_h$ in terms of $S = P + j Q$ and $\Sbar = P - j Q$ directly, we obtain the remarkable class of matrix dynamical systems: 
    \begin{align}
        \dot K_{hm} &= (\eta_h + \etabar_m) K_{hm} \nonumber \\
        \eta_h &= g^v_h\left(\sum_m K_{hm}, \sum_m \Kbar_{hm}, K_{hh}, x_h\right) \nonumber \\
        \dot x_h &= g^{x_c}_h\left(\sum_m K_{hm}, \sum_m \Kbar_{hm}, K_{hh}, x_h\right)\label{eq:minimalist_coupling_dynamics_system}
    \end{align}

\begin{figure}[H]
        \centering
        \begin{subfigure}[b]{0.32\textwidth}
            \centering
            \includegraphics[width=\textwidth]{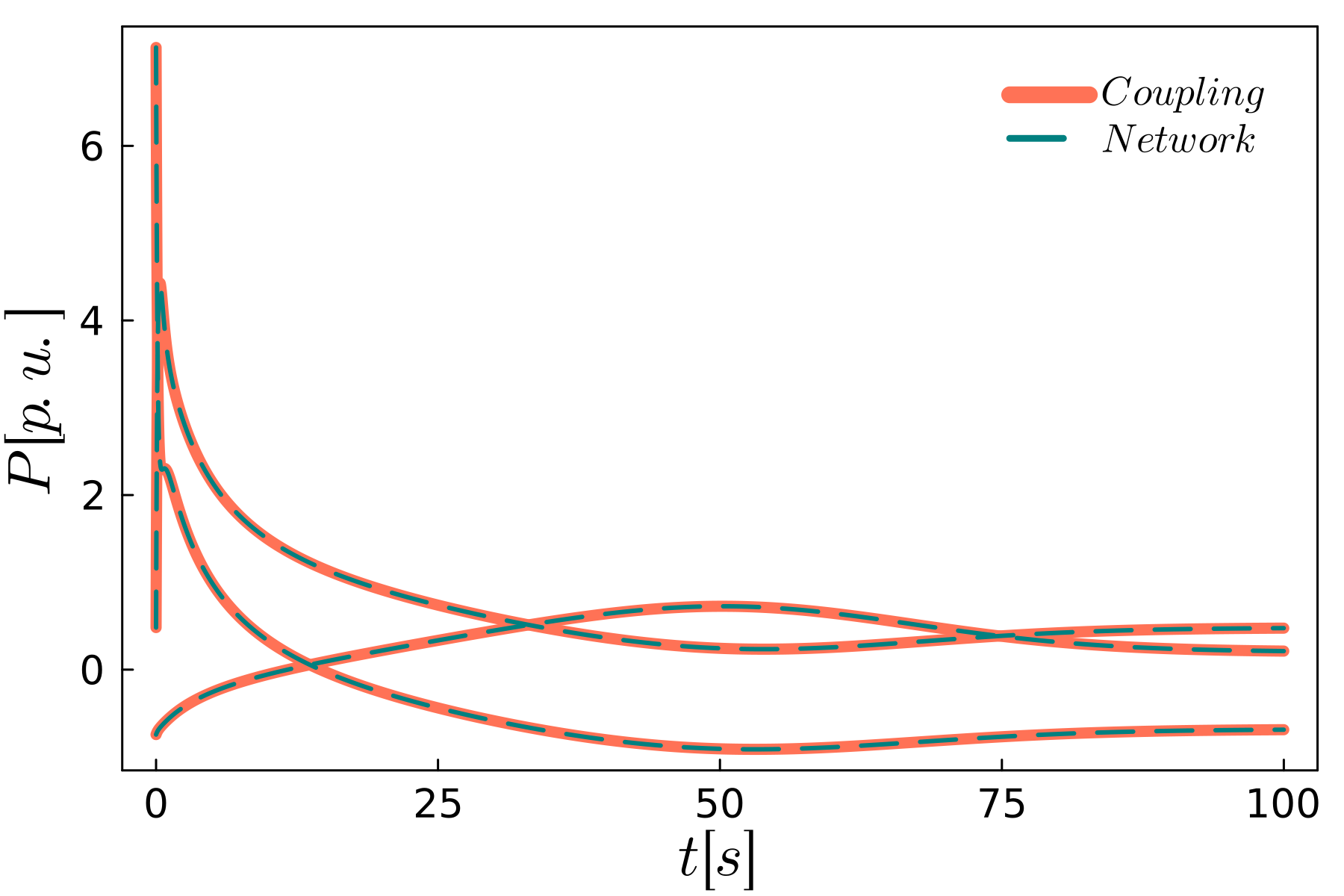}
        \end{subfigure}
        \hfill
        \begin{subfigure}[b]{0.32\textwidth}
            \centering
            \includegraphics[width=\textwidth]{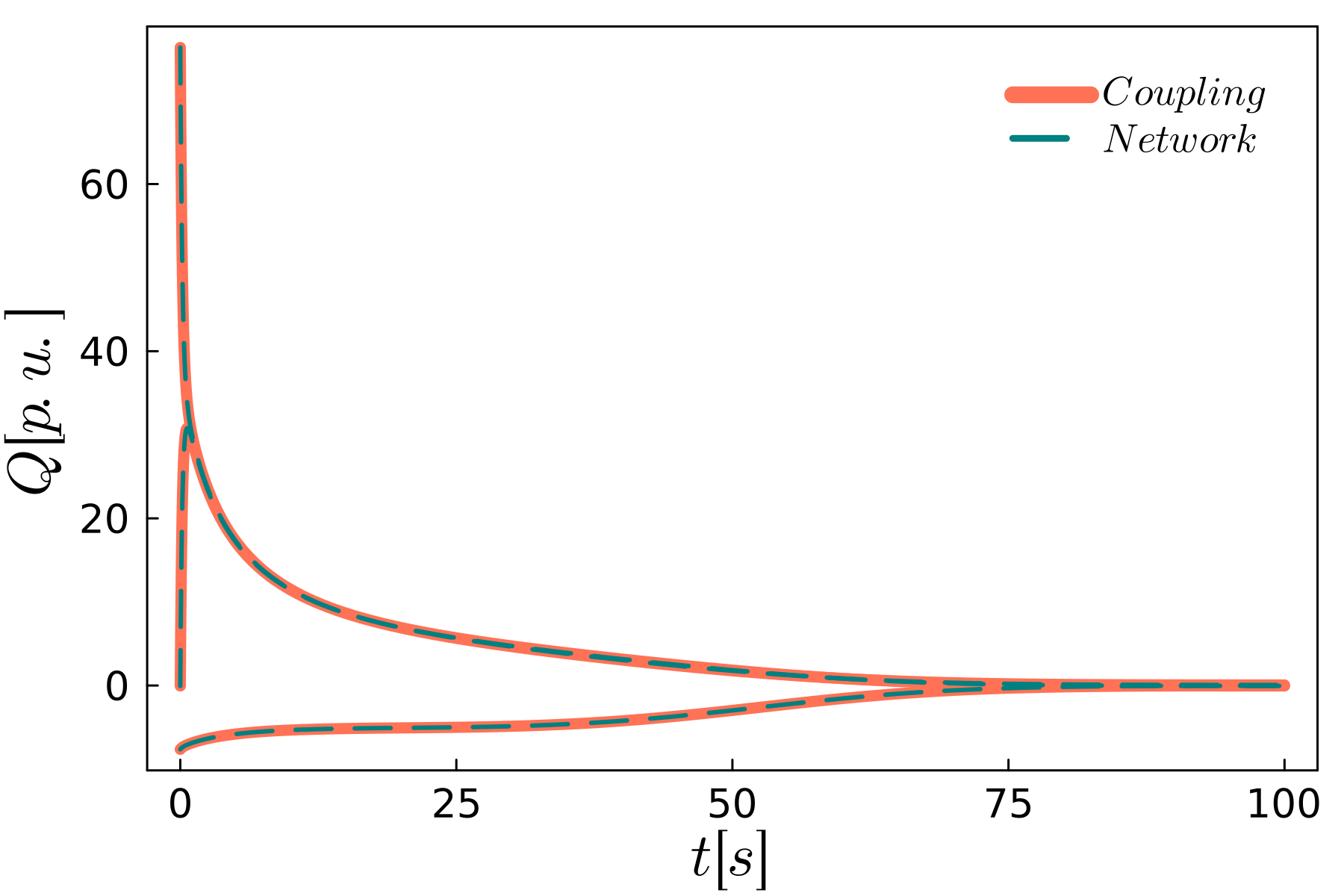}
        \end{subfigure}
          \hfill
        \begin{subfigure}[b]{0.32\textwidth}
            \centering
            \includegraphics[width=\textwidth]{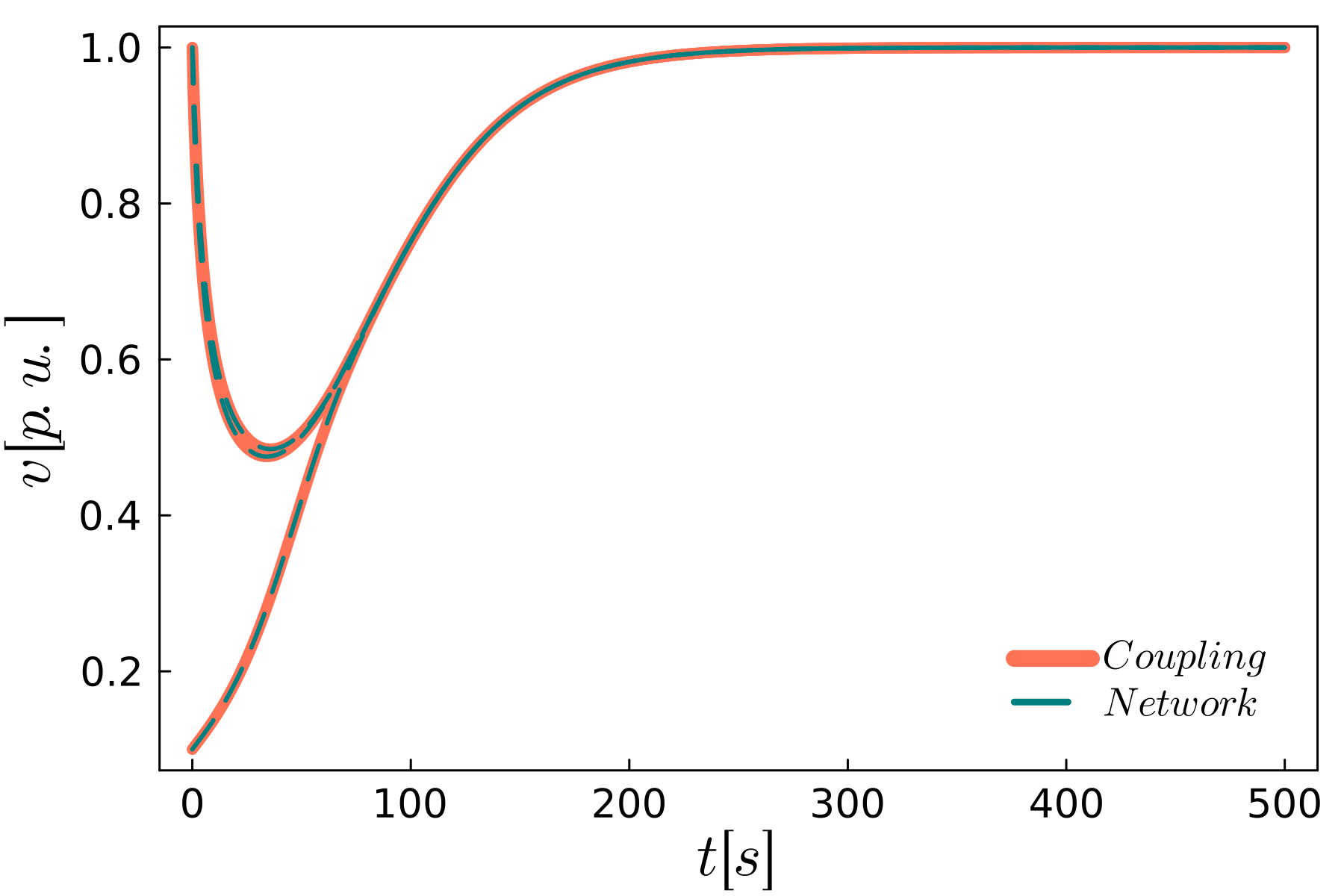}
        \end{subfigure}
        \caption{Comparison between the power and voltage given by the dynamics of a full network and the coupling dynamics given by equation \eqref{eq:coupling_dynamics}. It can be seen that both the coupling dynamics and the full network give the same results.}
        \label{fig:coupling_network_comparison}
    \end{figure}

    To illustrate the equivalence of the self-contained dynamical system, given by \eqref{eq:minimalist_coupling_dynamics_system}, to the dynamics of a full power grid we performed a simulation example. Using the algorithm introduced in \cite{buettner_synthetic_power_grids_2023} we generated a 3-bus power grid consisting of normal forms as the bus model. Figure \ref{fig:coupling_network_comparison} shows a comparison between the trajectories $P, Q$ and $\nu$ of the test grid and the equivalent coupling system \eqref{eq:minimalist_coupling_dynamics_system}. Using equations \eqref{eq:nu_coupling} and \eqref{eq:power_coupling} we calculate the voltage magnitude $v$ and the active and reactive power from the coupling terms. From figure \ref{fig:coupling_network_comparison} we can find that the full power grid and the coupling dynamics give the same results for the entire trajectory.

    The possibility of expressing power networks via these elemental coupling dynamics provides many opportunities for future research, e.g. for stability analyses of the collective dynamics of future power grids. In the following, we will use the coupling terms to design complex frequency dynamics that stabilize the power grid.
    
    \section{GFI Design as a Bilinear Control Problem}
    \label{sec:control_problems}

    In \cite{kogler_normalform_2022} the normal form is introduced as a way to parametrize the space of plausible grid actors. This allows us to obtain a unified description of inverter designs and existing machines. However, it is important to also ask which points in this space would provide stable dynamical laws for the power grid. 
    We assume that we have a perfect voltage source that we can steer freely, an assumption that is typical in the study of grid-forming control \cite{schiffer_droop_2014, colombino_dvoc_2017}. Below, we will see that the complex coupling variables allow us to cast this problem into the form of a bilinear-quadratic control problem. We also derive a quasi-local Lyapunov-function-based controller that is globally synchronizing.
    
    \subsection{Grid Forming Control as a Normal Form Tracking Problem}
\begin{figure}[H]
        \centering
        \includegraphics[width = 0.99 \textwidth]{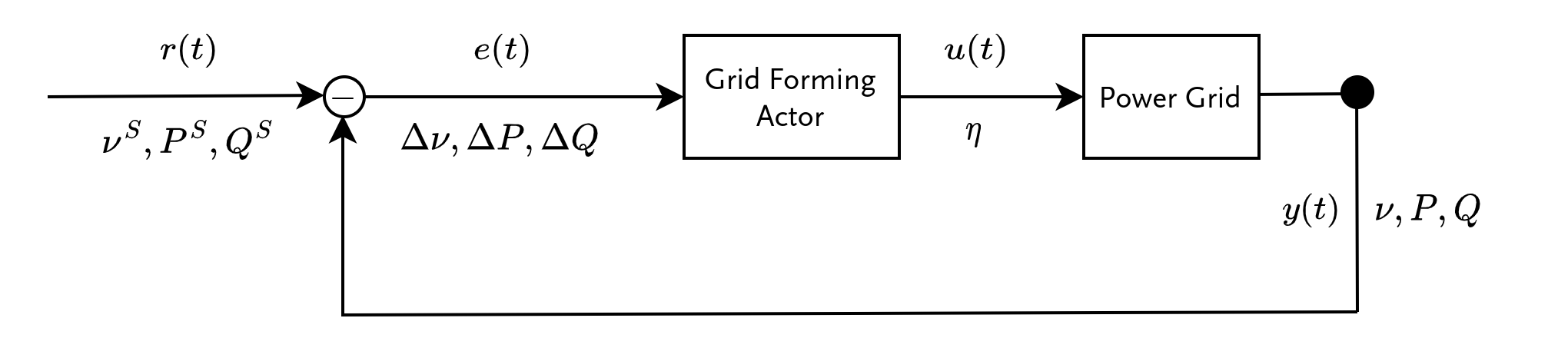}
        \caption{A block diagram that represents a single grid forming actor coupled to the power grid. The GFI prescribes a control input $u$, the complex frequency, to the power grid, which then leads to an observed output $y(t)$, the power flow and voltage, which is used as control feedback by the normal form.}
        \label{fig:normal-form}
    \end{figure}

    As noted above, the variables used in the normal form to express the coupling of the grid into the grid-forming actor are exactly those that a grid-forming actor seeks to control. We can assume that the grid-forming actor has (possibly time-dependent) set-points $\nu^s, P^s$, $Q^s$ for the desired active and reactive power and voltage amplitude square. Then we can write the normal form in terms of the error coordinates $e(t)$, the difference between the desired and actual quantities. The control problem of grid-forming actors is thus reduced to a \emph{tracking problem} with a feedback loop. The set-points act as the reference $r(t)$ which should be tracked, and the complex frequency is the control input $u(t)$ by which the power flow on the grid should be controlled. This is illustrated in \ref{fig:normal-form}. 
    If we require the feedback loop to be a linear time-invariant system, we arrive exactly at the internally linearized normal form equations already discussed in \cite{kogler_normalform_2022}, that are valid near the desired state.
    
    \subsection{Complex Voltage}
    \label{sec:bilinear_voltage}
    Viewed this way, the challenge of designing a good grid forming control is a decentral bilinear feedback control problem. If we use the complex voltages $v(t)$ as the states $y(t)$ and the complex frequency as the control input $u(t)$, the underlying bilinear system is reads as:
    \begin{align}
        \dot{y}(t) = \sum_{a=1}^A \eta_a(t) F^a \cdot y(t) 
    \end{align}
    where each system matrix $F^a$ has only one entry. The rows and columns of $F^a$ are the indices of the nodes in $\mathcal{N}$. The elements of $F^a$ are given by: 
    \begin{align}
    F^a_{hm} = 
        \begin{cases}
            1 & \text{if } a = k = m \\
            0 & \text{otherwise}
        \end{cases}.
    \end{align}
    where $h, m$ stands for the nodes in $\mathcal{N}$. This system has the advantage that it is very simple and elegant and is directly formulated in physically meaningful variables. However, the observed power and voltage amplitude mismatches that are required for the feedback are quadratic functions of the state variables, and the asymptotic state that should be achieved is not a fixed point but a limit cycle. The complex coupling formulation solves both of these issues.

    \subsection{Complex Couplings}
    The power grid dynamics are fully described by the coupling terms $K_{hm}$. We consider only those lines $l$ whose admittances are non-zero, corresponding to the set of links $\mathcal{L}$ (including self-loops) in the underlying power grid. We will treat $K_{hm}$, $\Kbar_{hm}$, $K_{mh}$ and $\Kbar_{mh}$ as separate variables, and the two directions of the edge between $h$ and $m \neq h$ as two distinct links $l = (h,m) \neq (m,h) = l'$. Thus the state vector $y$ is given by: 
    \begin{align}
        y(t) = (K_{l_1}, K_{l'_1}, ..., K_{l_d}, K_{l'_d}, \Kbar_{l_1}, \Kbar_{l'_1}, ..., \Kbar_{l_d}, \Kbar_{l'_d})^T.
    \end{align}
    The nodal complex frequencies $u(t) = (\eta_1, \etabar_1, .., \eta_n, \etabar_n)^T$ act as the control input into the system.
    
    We define the system matrices $R_a$ and $\Tilde{R}_a$ to bring the system into a bilinear form \ref{eq:bilinear_system}. We have two controls per node $h$: $\eta_h$, $\etabar_h$. The system matrices can be decomposed into the following blocks:

    \begin{align}
        R_a &= \left(
        \begin{array}{ c | c }
            O^a & 0 \\
            \hline
            0 & T^a
            \end{array}\right) \\
        \tilde R_a &= \left(
        \begin{array}{ c | c }
            T^a & 0 \\
            \hline
            0 & O^a
            \end{array}\right). 
    \end{align}
    We introduce the shorthand notation $o(l)$ and $t(l)$ for the origin and target of a link:
    \begin{align}
        l = (o(l), t(l)) \label{eq:links}
    \end{align}
    Then the origin matrix $O^a$ projects onto those links whose origin is $a$ i.e. $o(e) = a$. The rows and column indices of $O^a$ correspond to links of the graph, and we can write explicitly:
    
    \begin{align}
    O^a_{ll'} = \begin{cases}
        1 & \text{if } l = l' \text{ and } o(l) = a\\
        0 & \text{otherwise}
    \end{cases}
    \end{align}
    where $l,l'$ are the links in $\mathcal{L}$.
    
    The target matrix $T^a$ projects onto those links whose target is $h$:
    \begin{align}
    T^a_{ll'} = \begin{cases}
        1 & \text{if } l = l' \text{ and } t(l) = a \\ 
        0 & \text{otherwise}
    \end{cases}
    \end{align}
    With these matrices, we have the following bilinear form of the control problem:
    \begin{align}
        \dot{y}(t) = \sum_{h \in \mathcal N} \eta_a R_a \cdot y(t) + \etabar_a \Tilde{R}_a \cdot y(t) \; .
    \end{align}

    Taking these equations as the underlying system to be controlled by the complex frequency means that the tracking errors are now just linear combinations of the system state and reference signal. Thus, we have cast the challenge of finding stabilizing grid-forming controllers for the power grid, in terms of a decentral linear feedback control synthesis of a bilinear tracking control problem \eqref{eq:bilinear_system}.

    We should note that this formulation is not without challenges. The fact that we have one coupling per edge, while we have one voltage per node, implies that the complex frequency can not achieve arbitrary couplings from any starting position.
    Our linear state space decomposes into reachable layers. As we will see further below, this complicates the synthesis of elegant dynamical laws in this formulation.

    Furthermore, we want to note that the dynamics for the power flow \eqref{eq:s_line_dynamics} and the voltage square \eqref{eq:nu_dynamics} have a self-contained bilinear structure as well. The derivation can be found in the appendix \ref{sec:pf_bilinear}.
    
    \subsection{A quasi-local, error-minimizing and globally synchronizing grid control}
    So far, we have formulated a bilinear tracking control problem with linear feedback. The next problem to solve is to design a controller that effectively stabilizes the system. It is natural to consider a quadratic cost function in the error coordinates:
    \begin{align}
        V(S, \Sbar, \nu) &= \sum_h |S_h - S^s_h|^2 + (\nu_h - \nu_h^s)^2 \label{eq:square_norm_e}\\
        &= \sum_h |\Delta S_h|^2 + (\Delta \nu_h)^2 \geq 0 \nonumber.
    \end{align}

    This cost function can serve as a Control-Lyapunov function (CLF) for our system. CLFs are an extension of the Lyapunov function from general dynamical systems to systems with control inputs $u$ \cite{isidori_nonlinear_control_systems_1995}. According to Artstein's theorem \cite{sontag_nonlinear_stabilization_1989} if a CLF for a system exists then the system is asymptotically stabilizable, meaning that for any state $y$ a control $u(y)$ can be constructed that asymptotically guides the system back to a fixed point.

    Recall that a control-affine system is given by
    \begin{align}
        \mathbf{\dot{y}} = \mathbf{F(y)} + \mathbf{G(y) u}.\tag{\ref{eq:control_affine}}
    \end{align}

    For such systems, a stabilizing control can be explicitly constructed from any suitable $V$ for which the gradient $V$ does not vanish except at the fixed point. The most well-known formula for constructing a stabilizing controller is given by Sontag's formula \cite{sontag_nonlinear_stabilization_1989}. However, in our context we have $\mathbf{F(y)} = 0$, and a stabilizing control is already given by:
     \begin{align}
        u^{c} = - \left( \nabla V \cdot \mathbf{G(y)} \right)^T \label{eq:control_easy} \;,
    \end{align} 
    where $u^c$ is the control input, leading to the dynamics $\dot V = - u^{c} \dot u^{c}$.

    Using the square norm of the error coordinates of the tracking problem \ref{eq:square_norm_e} in this formula leads to the quasi-local control law:
    \begin{align}
        - \delta \eta_{h}^{c} := \Delta S_h \Sbar_h + 2 \Delta \nu_h \nu_h + \sum_m K_{mh} \Delta \Sbar_m.
    \end{align}
    The full derivation of this control law, as well as a more in-depth introduction to CLF, can be found in the appendix \ref{sec:app_CLF}.

    The control law \eqref{eq:control_easy} has the property that:
    \begin{align}
    \dot V = - \sum_h \delta \etabar_{h}^{c} \delta \eta_{h}^{c} \leq 0. \label{eq:eta_control}
    \end{align}

    Thus, the mismatch between the realized grid state and the desired grid state can not grow under this dynamical law. As the cost function is positive definite and vanishes at the desired grid state, we immediately obtain that this control law also stabilizes the system's operating state. However, we can not conclude that the system will always find the global minimum of the cost function. While the cost function is quadratic and thus convex on the full space of couplings, its projection on the reachable layers is not. Therefore, we have to contend with local minima of the cost function. However, these local minima are still globally synchronized because they still imply that $\dot V$ and thus $\delta \eta$ go to zero. We show in Appendix~\ref{sec:voltage_existence} that the local minima of $V$, subject to the constraint that the $h$ arise from some $v$, exactly require $\eta_{h}^{c} = 0$.
    
    It is also noteworthy that this control law seems to stabilize fair power-sharing at the global minimum of the cost function, without coupling the power imbalance to the frequency (see Section~\ref{sec:LCF-example}). The system always synchronizes exactly at the design frequency.

    All points that satisfy the power flow equations have $V = 0$ and thus are stable. This is in direct contrast to the requirements typically imposed on the power flow solution for guaranteeing stability. For example, \cite{he_complex_frequency_sync_2022} require that the difference between the phase angles between voltages is smaller, and in particular, not larger than $\pi/2$. Thus $v_i \vbar_j$ has to have a positive real part. Stability of the Kuramoto model is likewise only guaranteed if the effective network Laplacian, weighted with the cosine of the phase angle differences, is positive definite \cite{witthaut2022collective}. This follows easily from the condition that the differences are smaller than $\pi/2$ if the system is lossless. However, the condition is complicated in general.

    The disadvantage of this controller is that it is not decentral but only quasi-local as $\eta^{c}_h$ requires information about the state of the lines and nodes adjacent to $h$. Thus, the resulting controller is not in the space of behaviors parametrized by the normal form \cite{kogler_normalform_2022}. It is, however, similar in its communication needs to distributed averaging controllers proposed to tackle secondary control objectives in power grids\cite{andreasson2012distributed, schiffer2017robustness}. It should also be noted that this controller minimizes the cost function, but is not an optimal control in the sense that it minimizes the integral of the cost over time.
    
    We consider this controller and the intriguing dynamics it provides as a proof of concept that the formulations given here are highly promising for further analytical investigation.
    
    It will be especially worthwhile to investigate how to incorporate a decentrality constraint. Furthermore, it has to be noted that the controller highly depends on the choice of the cost function to use as CLF. Hence, it is valuable to find and analyze other possible CLFs and their corresponding controllers. This is beyond the scope of this work and will be addressed in subsequent papers. 

    \subsubsection{Examples}\label{sec:LCF-example}
\begin{figure}[H]
        \centering
        \begin{subfigure}[b]{0.32\textwidth}
            \centering
            \includegraphics[width=\textwidth]{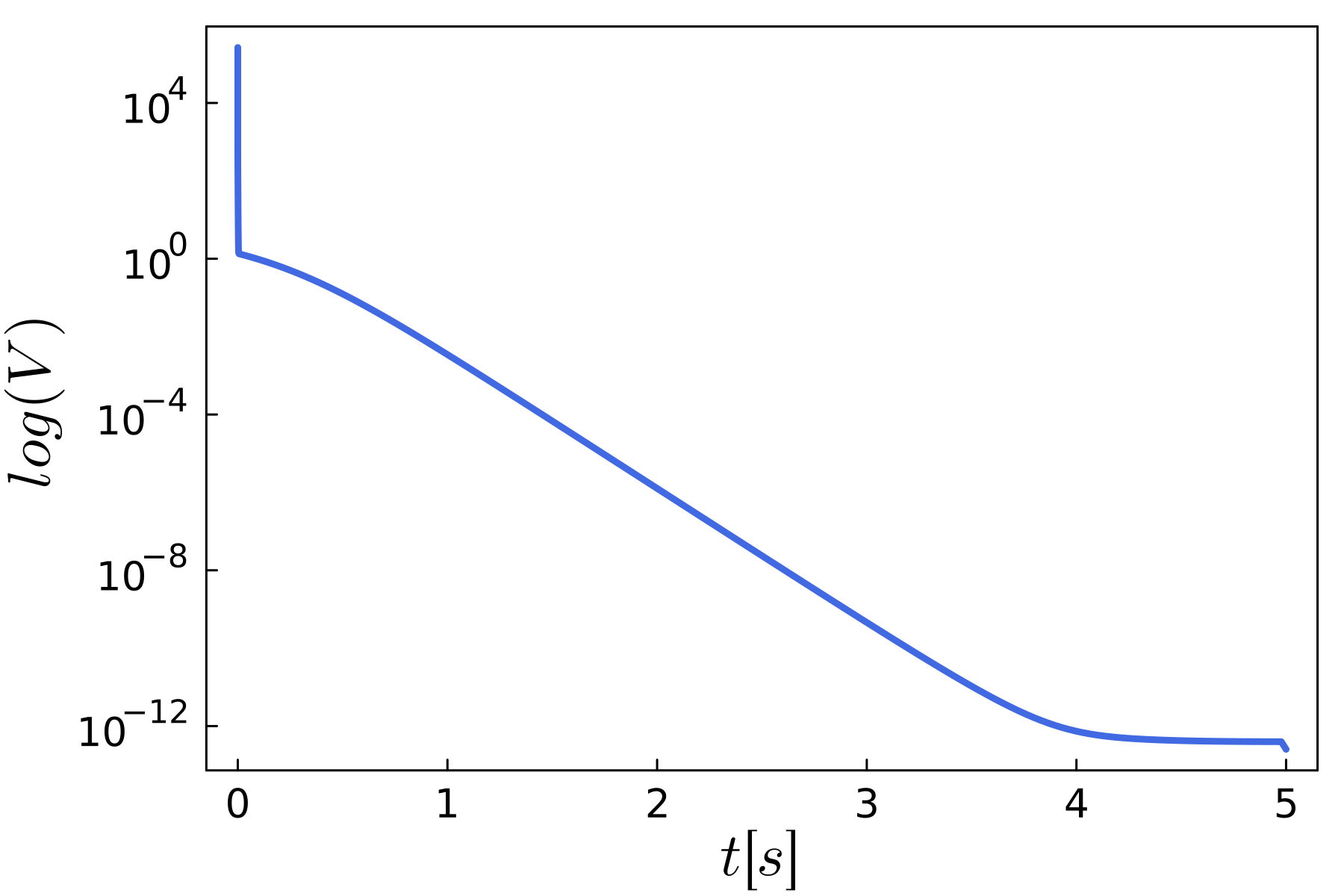}
        \end{subfigure}
        \hfill
        \begin{subfigure}[b]{0.32\textwidth}
            \centering
            \includegraphics[width=\textwidth]{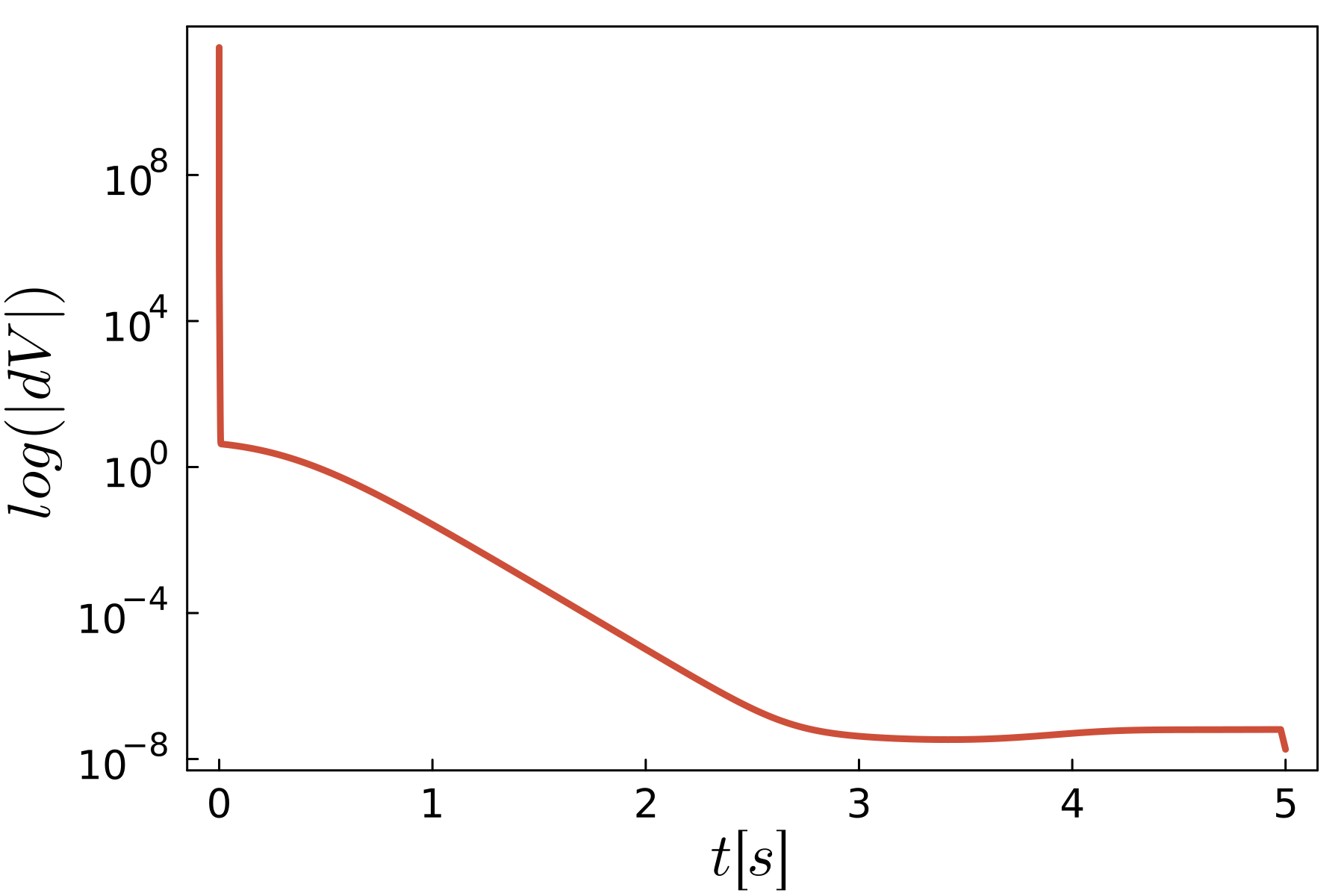}
        \end{subfigure}
        \hfill
        \begin{subfigure}[b]{0.32\textwidth}
            \centering
            \includegraphics[width=\textwidth]{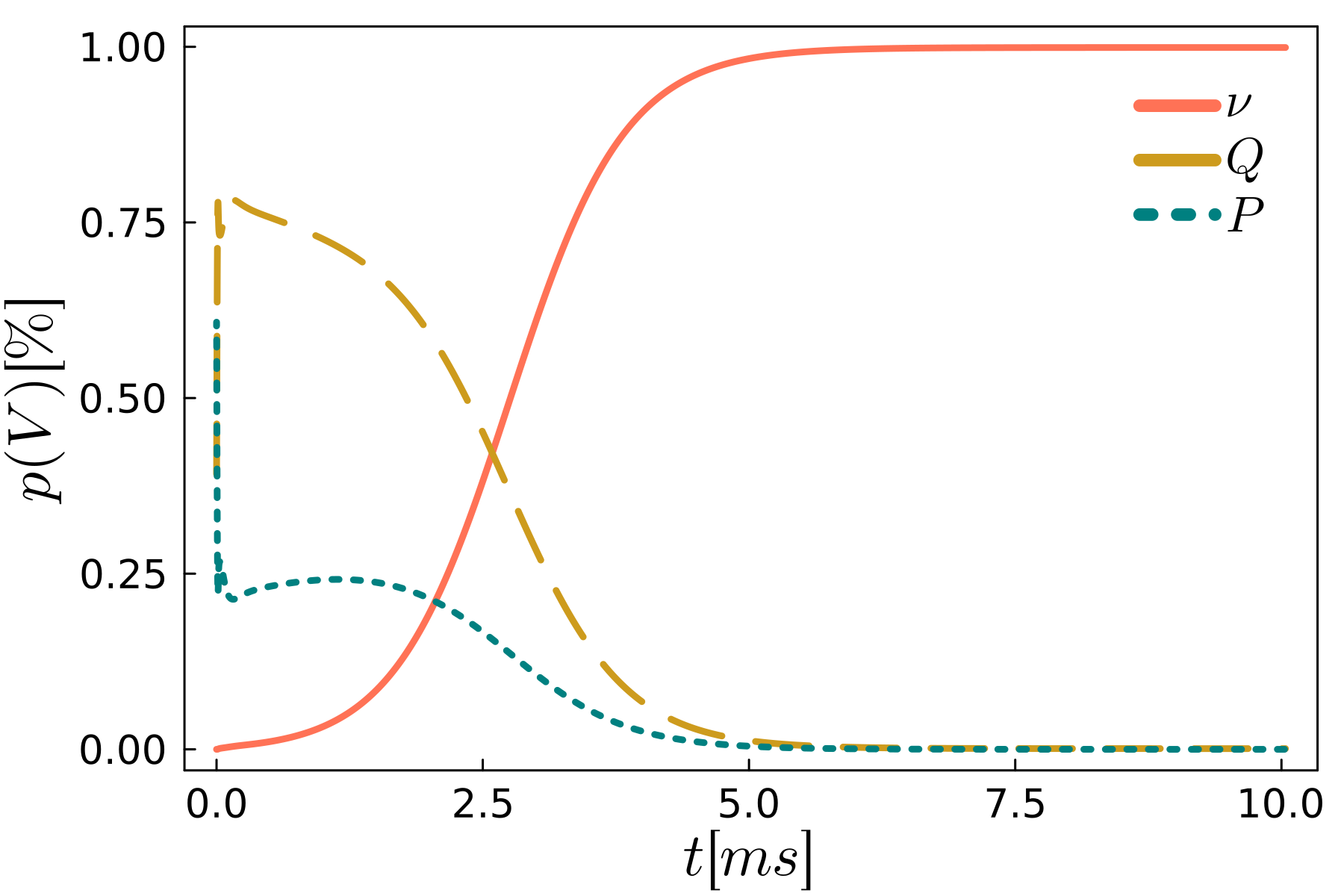}
        \end{subfigure}
        \caption{The CLF $V$ and its derivative $dV$ after a voltage perturbation to a node. It can be seen that the CLF reduces rapidly in the first milliseconds by compensating the active and reactive power errors, which can be seen in the right-hand figure which shows the relative shares of the CLF to its three components $\Delta P, \Delta Q, \Delta \nu$. After the initial drop, the CLF decreases further and then saturates after approximately 4 seconds.}
        \label{fig:V_dV}
    \end{figure}

    Using the same algorithm as in section \ref{sec:bilinear_coupling}, we generate a 10-bus synthetic power grid to validate the control law in a simulation example. We implement the complex frequency $\eta^{c}$ given by the control law \eqref{eq:eta_control}. We calculate the voltage transients according to equation \ref{eq:voltage_dynamics} and then define the currents and powers using Ohm's law and Kirchhoff's law.
    As an example, we decrease the voltage at a bus to 0.5 [p.u.] and study the CLF $V$ and its derivative $dV$ as shown in figure \ref{fig:V_dV}. We can see that the CLF is rapidly decreasing in the first milliseconds of the simulation. Furthermore, we also analyze which of the error coordinates $\Delta P, \Delta Q, \Delta \nu$, gives the largest share to the CLF. From the relative shares, we find that this rapid decrease is due to the reduction of the error in the active and reactive power. After this first initial drop, the CLF decreases further, which corresponds to the reduction in the voltage magnitude errors. Then the CLF saturates at $10^{-17}$ after approximately 4 seconds as also reflected by its derivative which reaches $10^{-8}$ after approximately 3 seconds.

    Additionally, we use the control law for a proper tracking problem. In this example, we simulate a random superposition of sinusoidal fluctuations of the active power set-points at all nodes. Each fluctuation has a different period and amplitude and runs for one respective period. As these are drawn independently the resulting set points are inconsistent during the fluctuation.
    
    Note that due to the non-zero derivatives of the set points, we do not have the condition that $\dot{V}< 0$. As soon as the variation of the set-points stops, $V$ has to decrease. Nevertheless, the dynamics is constantly working to minimize the cost function and successfully tracks the desired set points.
    
\begin{figure}[H]
            \centering
            \begin{subfigure}[b]{0.32\textwidth}
                \centering
                \includegraphics[width=\textwidth]{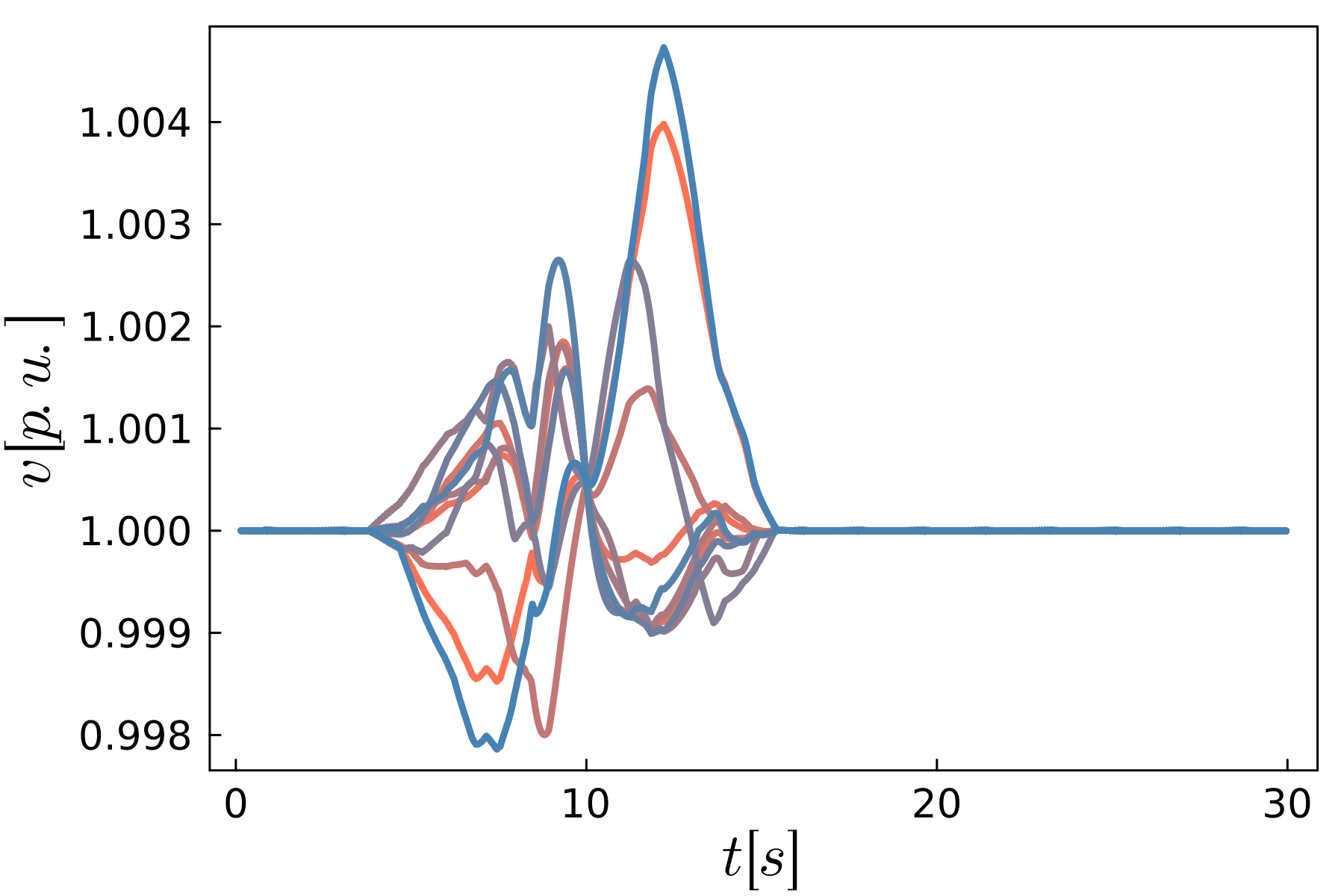}
            \end{subfigure}
            \hfill
            \begin{subfigure}[b]{0.32\textwidth}
                \centering
                \includegraphics[width=\textwidth]{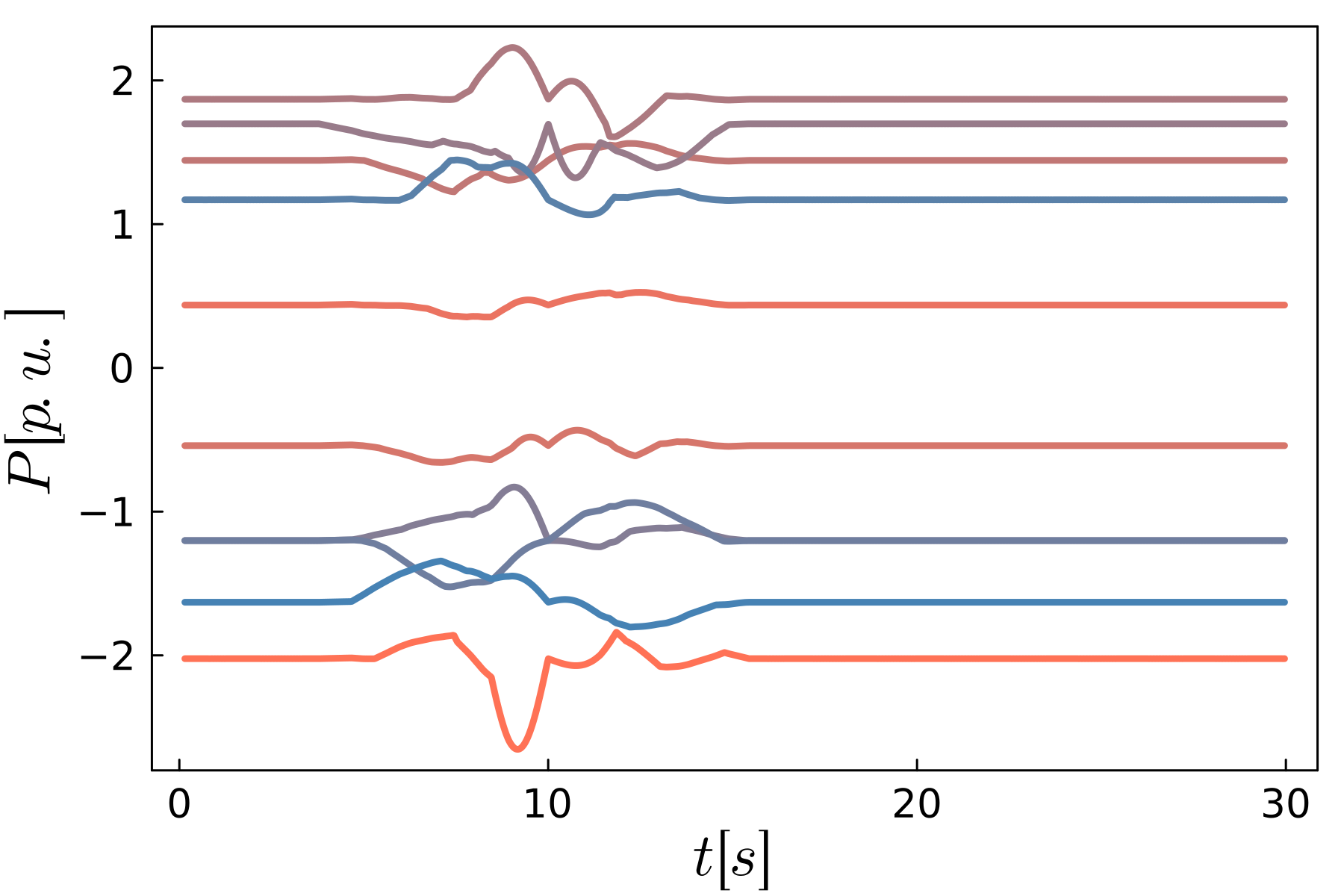}
            \end{subfigure}
            \hfill
            \begin{subfigure}[b]{0.32\textwidth}
                \centering
                \includegraphics[width=\textwidth]{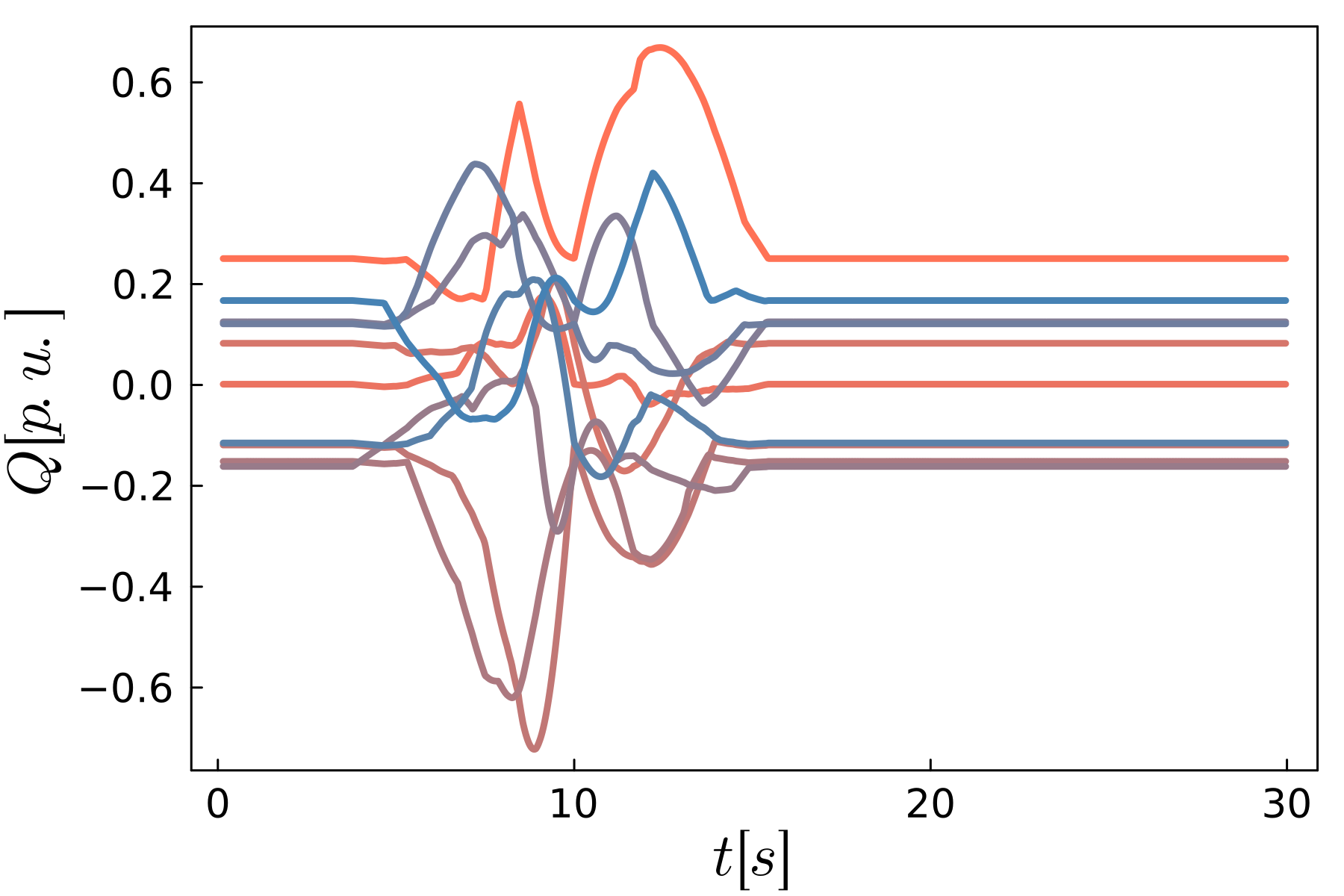}
            \end{subfigure}
            \caption{Voltage magnitude, active and reactive power transients during the set-point fluctuations of the 10-node test system.}
            \label{fig:tracking}
        \end{figure}

\begin{figure}[H]
            \centering
            \begin{subfigure}[b]{0.49\textwidth}
                \centering
                \includegraphics[width=\textwidth]{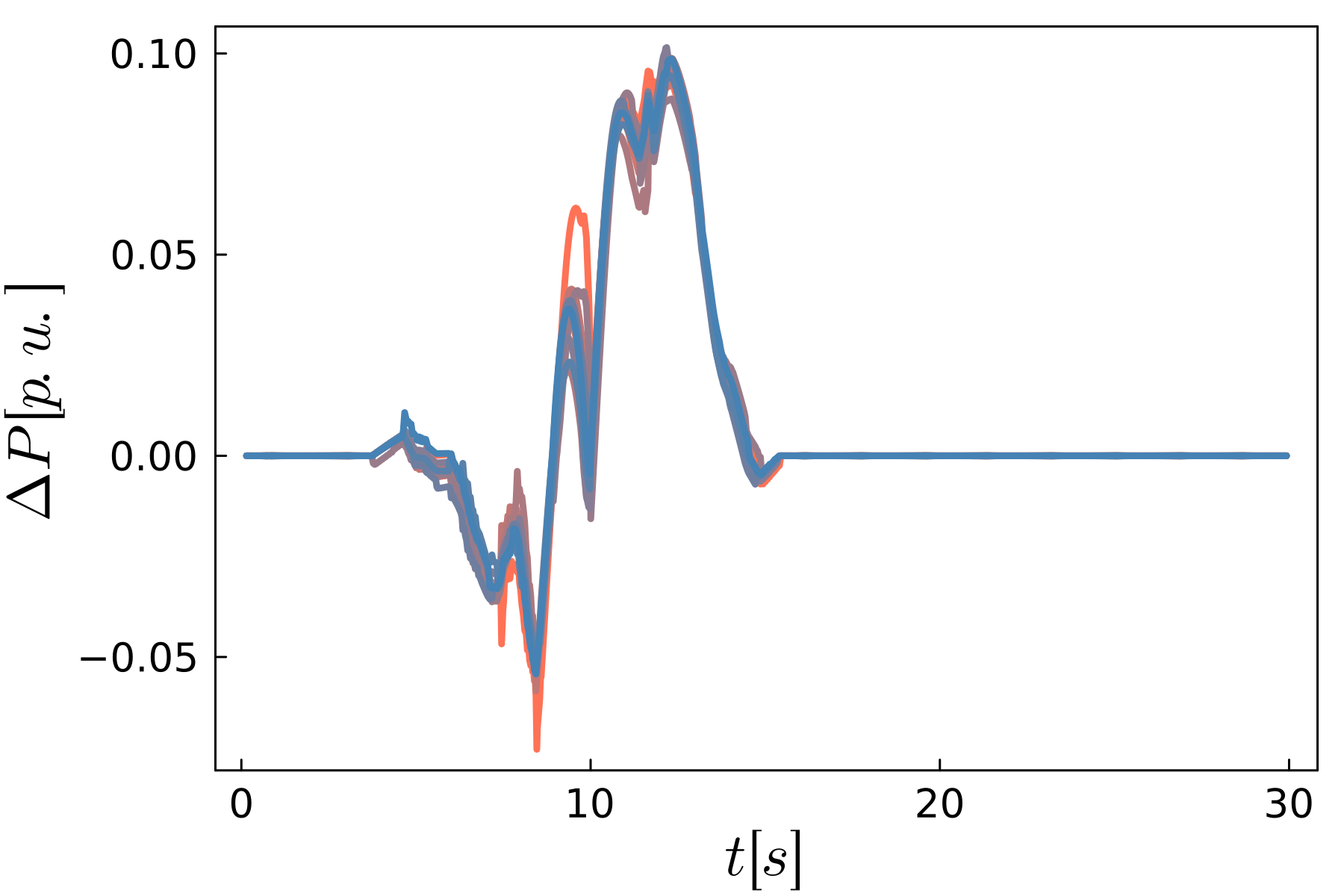}
            \end{subfigure}
            \hfill
            \begin{subfigure}[b]{0.49\textwidth}
                \centering
                \includegraphics[width=\textwidth]{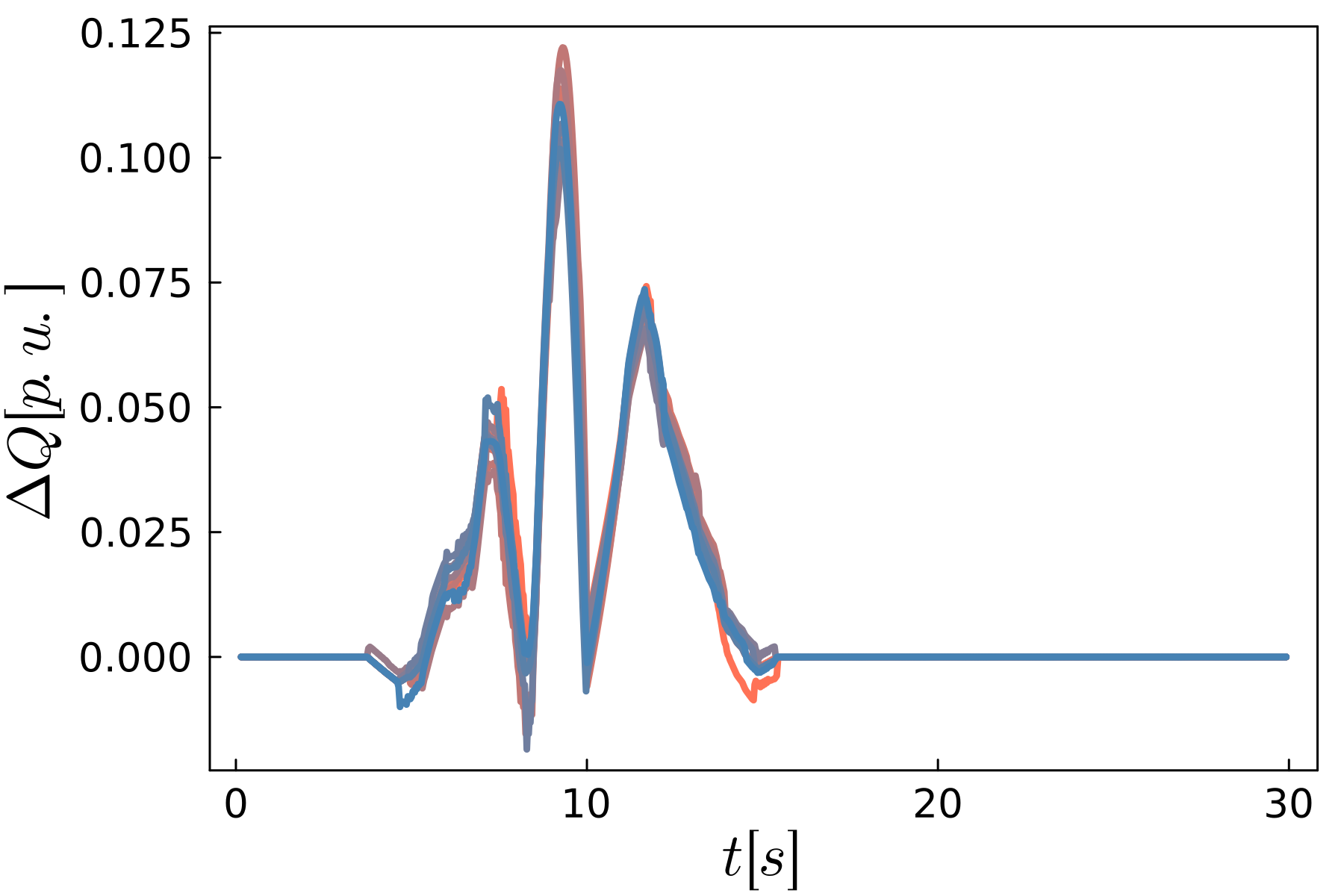}
            \end{subfigure}
            \caption{Errors of the active and reactive power transients during the set-point fluctuations of the 10-node test system.}
            \label{fig:tracking_delta}
        \end{figure}

    Figure \ref{fig:tracking} provides the transients of the voltage magnitude and the active and reactive power. We find that the active and reactive powers are adjusted to the new set points. The system does not become unstable, although the set points are highly unbalanced during the transient period. Remarkably, the voltage magnitudes stay very close to their nominal values, although the system is under severe stress. This is a further success of the designed control law. Figure \ref{fig:tracking_delta} shows the errors of the rated power and the set points. As the active power set points are highly unbalanced during the fluctuations, it is impossible for the system to perfectly follow the set points. Instead, our control law adjusts the power outputs such that all nodes share the mismatch, which is also referred to as equal sharing. After all fluctuations have vanished, the control law smoothly brings the system back into the previously synchronized and stable state. 

    Further examples, such as the so-called black start capabilities and simulations for larger networks, can be found in the appendix \ref{app:lcf_examples}.
    
\section{Conclusion}
    The stability of future power grids, particularly in the context of the wide-scale integration of renewable energy sources via power-electronic inverters, remains a challenging topic. While there have been numerous stability analyses of power-electronic inverters connected to an infinite bus bar \cite{schiffer_droop_2014, seo_dvoc_2019}, and stability analysis of networks consisting purely of conventional generation \cite{witthaut2022collective, auer2016impact}, the stability of inverter-dominated networks is not well understood and is an area of active research.
    
    To address these challenges, the normal form \cite{kogler_normalform_2022}, a dynamic technology-neutral model for grid-forming actors, has recently been introduced. The normal form can describe the dynamics of all grid-forming actors, including grid-forming inverters and conventional synchronous generators, and can thus capture the transitional periods when a reduced number of synchronous generators are still connected to the grid, as well as future highly heterogeneous inverter-dominated systems. 
    
    In this paper, we have shown that the concept of the complex frequency \cite{milano_complex_frequency_2022} combined with newly introduced dynamic complex couplings \eqref{eq:coupling_dynamics}, allows for an elegant adaptive network formulation of the Kuramoto model with inertia. When combined with the normal form description of grid-forming actors, we obtain a self-contained matrix dynamics for general power grid dynamics in which the grid topology does not explicitly appear. Beyond what we have shown here, the complex coupling dynamics allow for rewriting a large class of dynamical equations as adaptive.
    
    As considerable progress has been made recently in understanding the properties of adaptive networks \cite{berner2023adaptive, Sawicki_adaptivity_2023}, this opens up a new avenue for understanding power grids beyond the currently used generator models. We expect that our formulations will be useful in deriving widely applicable analytic stability results in the future.
    
    Furthermore, we have seen that the design of stable grid-forming actors can be cast as a bilinear control problem. The control problem simplifies due to the control-affine nature of the bilinear system and allows us to define a stabilizing control input. By defining a suitable Control-Lyapunov function, a control law has been derived that leads to global synchronization. However, it should be noted that the resulting controller is not completely decentralized, as it requires information from adjacent nodes. Thus it resembles distributed averaging control \cite{andreasson2012distributed} which is usually considered for secondary control objectives, such as restoring frequency and power balance. Both of these are achieved by our controller as well.
    
    The formulation of power grids using the combination of complex frequency and normal form shows surprising elegance. This paper has mapped out some, though by no means all, of the relevant results from control theory, electrical engineering, and statistical physics to demonstrate that there is considerable synergy in this approach. Excitingly, the methods include the full physics of balanced three-phase power grids, including losses and generic voltage dynamics. These are known to be important to obtain a full understanding of the collective dynamics of the system \cite{hellmann_2020_lossy_coupling, auer2016impact}. The proposed models and formulations thus offer a pathway for further research and development in this area, to ensure the stability and reliability of future renewable energy systems.

    \section{Acknowledgements}
    The authors would like to thank Jakob Niehues and Jürgen Kurths for their discussions and detailed comments on the manuscript. The work was in parts supported by DFG Grant Number KU 837/39-2 and BMWK Grant Number 03EI1016A. Anna Büttner acknowledges support from the German Academic Scholarship Foundation.
    \bibliographystyle{iopart-num}
    \bibliography{main}
    \newpage


\appendix
\section*{Appendix}
    In Appendix~\ref{sec:app_commplex}, we collect useful information on complex variables for working with the formulation introduced in the main text. Appendix~\ref{sec:line-dynamics} gives a minor extension of the formulation above to the case of dynamical lines with a homogeneous ratio of resistance to inductance. Appendix~\ref{sec:app_CLF} gives further on the new control law given above, including a more detailed derivation and the characterization of the local minima of the control Lyapunov function.

    
    \section{Complex variables}\label{sec:app_commplex}

    The formulation of the power grid in terms of complex variables is succinct but requires that some care is taken, especially when working with derivatives of the functions involved. This is due to the fact, that complex conjugation of a complex number $z$ is not a complex differentiable function. We can work around this fact by treating $\zbar$ as a separate variable. To illustrate how this arises, we will start from a real-valued system and derive the correct equations. For this section, we will call the complex conjugate of a variable $z^*$ to differentiate it from $\zbar$ more clearly.
    
    Consider a dynamical system with two $d$-dimensional vectors $x$ and $y$.
    \begin{align}
         \dot x &= f^x(x,y)\\
         \dot y &= f^y(x,y)    
    \end{align}
    We can write $z = x + j y$. Then we can write the dynamics as:
    \begin{align}
         \dot z &= f^z(z)\\ &:= f^x(\Re(z),\Im(z)) + j f^y(\Re(z),\Im(z))
    \end{align}
    unless $f^x$ and $f^y$ satisfy the Cauchy-Riemann equation, $f^z(z)$ is not differentiable in the complex sense as a function of $z$, even if $f^x$ and $f^y$ are real differentiable. In particular, $\Re(z)$ is not a differentiable function of $z$.
    
    However, if we introduce two new variables as ''complex linear combinations'' $z = x + j y$ and $\zbar = x - j y$, we can write $x = \frac{1}{2} (z + \zbar)$ which is separately differentiable in $z$ and $\zbar$. To take this derivative, we treat $z$ and $\zbar$ as independent complex variables. Taking this seriously means that $x$ and $y$ can now be complex. If we evaluate the system on $\zbar = z^*$, that is, we enforce by hand that $\zbar$ really is the complex conjugate, $x$ will be real.
    
    We can then cast the system as:    
    \begin{align}
        \dot z &= f^{hol}(z, \zbar)\\ &:= f^x(\frac12 (z + \zbar), \frac1{2j} (z - \zbar)) + j f^y(\frac12 (z + \zbar), \frac1{2j} (z - \zbar))\\
        \dot {\zbar} &= {\overline f}^{hol}(z, \zbar)\\ &:= f^x(\frac12 (z + \zbar), \frac1{2j} (z - \zbar)) - j f^y(\frac12 (z + \zbar), \frac1{2j} (z - \zbar))
    \end{align}
    
    Now $f^{hol}$ and ${\overline f}^{hol}$ have a good chance to be differentiable as functions of $z$ and $\zbar$. In particular, this will be the case whenever we can analytically extend $f^x$ and $f^y$ into the complex plane. For example, this will work for polynomials.
    
    \subsection{Wirtinger derivatives}
    
    While extending the function to the complex plane might not always be possible, we can still access the ''derivative in the $z$, $\zbar$ direction'' without actually extending the underlying real functions. Write $\partial_x$ for $\frac{\partial}{\partial x}$ then for the complexified function we have:

    \begin{align}
        \partial_z  f^{hol}(z, \zbar) &= \frac{1}{2} (\partial_x - j \partial_y) (f^x + j f^y)\\
        \partial_\zbar  f^{hol}(z, \zbar) &= \frac{1}{2} (\partial_x + j \partial_y) (f^x + j f^y)\\
        \partial_z \overline{f}^{hol}(z, \zbar) &= \frac{1}{2} (\partial_x - j \partial_y) (f^x - j f^y)\\
        \partial_\zbar \overline{f}^{hol}(z, \zbar) &= \frac{1}{2} (\partial_x + j \partial_y) (f^x - j f^y),
    \end{align}
    
    The right-hand side is well-defined without complexification. It can be evaluated using only the real functions $f^x$ and $f^y$ and their derivatives. The derivative operators $\partial_{z/\zbar} = \partial_x \mp j \partial_y$ applied to a complex-valued function on the real domain of $x$ and $y$ is called the Wirtinger derivative. Note that the Cauchy-Riemann equation is simply $\partial_\zbar f = 0$.

    Note that we have 
    \begin{align}
        \partial_z  z &= 1\\
        \partial_\zbar  z &= 0\\
        \partial_z \zbar &= 0\\
        \partial_\zbar \zbar &= 1,
    \end{align}
    as might be expected.
    
    \subsection{Extrema with Wirtinger derivatives}
    To illustrate the use of these equations, consider the optimization of a real function $V(x,y)$. The first order condition for extrema is 
    
    \begin{align}
    \partial_x V &= 0 \\
    \partial_y V &= 0
    \end{align}
    
    If we write the function in terms of $V(z, \zbar)$ we can write this condition as
    \begin{align}
    \partial_z V &= 0 \\
    \partial_\zbar V &= 0
    \end{align}

    Take for example $V = xy + x^2 + y^2$, then $V = z \zbar + \frac1{4j}(z + \zbar)(z - \zbar)$. The complex derivative conditions then give 

    \begin{align}
    \partial_z V &= \zbar + \frac1{4j}(z - \zbar + z + \zbar) \\
    0 &= \zbar + \frac1{2j}z \\
    \partial_\zbar V &= z + \frac1{4j}(z - \zbar - z - \zbar)\\
    0 &= z - \frac1{2j}\zbar
    \end{align}
    
    Which can readily be solved by inserting the second equation into the first. Note that the second equation can be obtained from complex conjugating the first, which can also be verified directly from the definition of the Wirtinger derivatives applied to a real function. Depending on the functions at hand, it can be considerably simpler or harder to work in complex coordinates.

    
    \section{Line Dynamics}
    \label{sec:line-dynamics}
    The main body of the paper is concerned with the dynamics of power systems and power lines in the quasi-steady state approximation, that is, the current on the lines is given by equation \eqref{eq:line-current}:
    \begin{align}
        i^{qss}_{hm} &= Y_{hm} (v_h - v_m) \; . \tag{\ref{eq:line-current}}
    \end{align}
    In reality, a change in voltage at one end of the line does not lead to an instantaneous change in current at the other. A dynamical model for lines that takes this into account better is the RL-line model. This describes lines as a series circuit of an inductance and a resistor $R$. The inductance is typically called $L$. To avoid confusion with the Laplacian matrix we call it $\Lambda$. With complex voltage and current and in a frame rotating with $\Omega$ this leads to the differential equation:

    \begin{align}
        \Lambda_{hm} \frac{di_{hm}}{dt} &= v_h - v_m - R_{hm} i_{hm} - j \Omega \Lambda_{hm} i_{hm} 
    \end{align}

    introducing the complex impedance $Z_{hm} = R_{hm} - j \Omega \Lambda_{hm}$, which is the reciprocal of the admittance $Y_{hm}$ we can write this as

    \begin{align}
        \Lambda_{hm} \frac{di_{hm}}{dt} &= v_h - v_m -  Z_{hm} i_{hm} \\
        \Lambda_{hm} \frac{di_{hm}}{dt} &= Z_{hm} (Y_{hm}(v_h - v_m) -  i_{hm}) \\        
        \frac{di_{hm}}{dt} &= \frac{Z_{hm}}{\Lambda_{hm}} (i_{hm}^{qss} -  i_{hm}) \\        
        \frac{di_{hm}}{dt} &= \left(\frac{R_{hm}}{\Lambda_{hm}} + j\Omega\right) (i_{hm}^{qss} -  i_{hm})
    \end{align}

    For the power on the line, we immediately obtain    

    \begin{align}
        \frac{dS_{hm}}{dt} = \eta_h S_{hm} + \left(\frac{R_{hm}}{\Lambda_{hm}} - j\Omega\right) (S_{hm}^{qss} -  S_{hm})
    \end{align}

    And $S_{hm}^{qss}$ follows the equation of Section~\ref{sec:pf_bilinear}.

    A remarkable simplification arises if $\frac{R_{hm}}{\Lambda_{hm}}$ is assumed to be homogeneous throughout the network. This is plausible as both $R$ and $\Lambda$ are proportional to the length of the line. Call their ratio $\alpha_{rl} = \frac{R}{\Lambda}$, then the dynamics of the power can be given in terms of the $K_{hm}$ directly:

    \begin{align}
        \dot S_{h} &= \eta_h S_{h} + \left(\alpha_{rl} - j\Omega\right) (S_{h}^{qss} -  S_{h}) \\
        \dot S_{h} &= \eta_h S_{h} + \left(\alpha_{rl} - j\Omega\right) \left(\sum_{m} K_{hm} -  S_{h}\right)
    \end{align}

    Thus, this model of dynamic lines leads to the same adaptive coupling formulation in terms of $K_{hm}$, the only difference is that the nodal dynamics get augmented by what is essentially a low-pass filter for the part of the nodal power variation that arises due to the coupling.

\section{Power Flow Variables}
    \label{sec:power_flow_vars}
    The complex power flow $S_{hm}$ on a line connecting node $h$ and $m$ is defined as:
    \begin{align}
        i_{hm} &= Y_{hm} (v_h - v_m)\\
        S_{hm} &= v_h \ibar_{hm} = \Ybar_{hm} \nu_h - \Ybar_{hm} v_{h} \vbar_m
    \end{align} 
   The dynamics for the power flow then lead to:
    \begin{align}
        \dot S_{hm} &= \dot v_h \Ybar_{hm} \vbar_h + v_h \Ybar_{hm} \dot \vbar_h - \dot v_h \Ybar_{hm} \vbar_m - v_h \Ybar_{hm} \dot \vbar_m\\
        &= \eta_h \Ybar_{hm} \nu_h + \etabar_h \nu_h \Ybar_{hm} - v_h \eta_h \Ybar_{hm} \vbar_m  - v_h \Ybar_{hm} \vbar_m \etabar_m\\
        &= S_{hm} \eta_h + \nu_h \Ybar_{hm} \etabar_h - v_h \Ybar_{hm} \vbar_m \etabar_m\\
        &= S_{hm} \eta_h + \nu_h \Ybar_{hm} \etabar_h  + S_{hm} \etabar_m - \nu_h \Ybar_{hm} \etabar_m \\
        &= S_{hm} (\eta_h + \etabar_m) + \nu_h \Ybar_{hm} (\etabar_h - \etabar_m). \label{eq:s_line_dynamics}
    \end{align}
    If we augment these with the dynamics for the voltage square $\nu_h = v_h \vbar_h$:
    \begin{align}
        \dot \nu_h &= \nu_h (\eta_h + \etabar_h). \label{eq:nu_dynamics}
    \end{align}
    This again provides a self-contained set of equations for the power flow and voltage amplitude as a function of the complex frequency at the nodes.
    
    \subsection{Power Flow Bilinear Structure} 
    \label{sec:pf_bilinear}
    For the power flow dynamics, the state vector $y$ is given by: 
    \begin{align}
         y(t) = (\nu_1, ..., \nu_n, S_{l_{1}}, S_{l’_{1}}, ..., S_{l_{d}}, S_{l'_{d}}, \Sbar_{l_{1}}, \Sbar_{{l'}_{1}}, ..., \Sbar_{l_{d}}, \Sbar_{{l'}_{d}})^T \; .
    \end{align}  
    $S_{l} = S_{mh}$ denotes the power flow on the link $l$ as seen from $t(l) = m$ and $S_{l’}$ denotes the flow in the opposite direction, as seen from $h$. 

    To bring the system into a bilinear form, we have to define system matrices. We have two controls per node $h$, $\eta_h$, $\etabar_h$, and we denote the corresponding system matrices as $N_a$ and $\tilde N_a$ respectively. The matrices can then be decomposed into the following block matrix form: 
    \begin{align}
        N_a &= \left(
        \begin{array}{ c | c | c }
            F^a & 0 & 0 \\
            \hline
            0 & O^a & 0 \\
            \hline
            M^a & 0 & T^a
            \end{array}\right) \\
        \tilde N_a &= \left(
        \begin{array}{ c | c | c}
            F^a & 0 & 0\\
            \hline
            M^{*a} & T^a & 0 \\
            \hline
            0 & 0 & O^a
            \end{array}\right). 
    \end{align}
    where the blocks $F^a$, $O^a$ and $T^a$ have been defined in sections \ref{sec:bilinear_voltage}-\ref{sec:bilinear_coupling}. $M^a$ is a mixed block handling the interaction between the voltages at the nodes and the power flow on the links and incorporates the line admittances. 
    
    The column indices of $M^a$ are links and the row indices are vertices. The elements of the mixed block are defined as:
    \begin{align}
        M^a_{lh} = 
        \begin{cases}
            Y_l & \text{if } k = a = o(l) \\
            -Y_l & \text{if } l = (h,  a)\\
            0 & \text{otherwise}
        \end{cases}
    \end{align}
    The block $\overline{M}_a$ is the element-wise complex conjugate of $M_a$. 










\section{Control Lyapunov Function }
    \label{sec:app_CLF}
    A function $V(y)$ is a CLF for a controlled dynamical system $\dot y = f(y,u)$ and the origin $0$, if: It is continuously differentiable; It is positive-definite, meaning that $V(y) > 0$ for all $y \neq 0$ and $V(0) = 0$; For all $y \neq 0$ there exists a $u$ such that
    \begin{align}
        \dot{V} < 0,
    \end{align}
    and there exists a $u^*$ such that $\dot{V}(0, u^*) = 0$.

    Control Lyapunov functions are particularly useful for control affine systems:
    \begin{align}
        \mathbf{\dot{y}} = \mathbf{F(y)} + \mathbf{G(y) u}.\tag{\ref{eq:control_affine}}
    \end{align}
    The derivative of a potential CLF $V$ for such a system is given by:
    \begin{align}
        \dot{V} = \frac{\partial V}{\partial \mathbf{y}}^T = \frac{\partial V}{\partial \mathbf{y}}^T \mathbf{f(y)} + \frac{\partial V}{\partial \mathbf{y}}^T \mathbf{G(y) u}  = \mathbf{a(y)} + \mathbf{b(y)}^T \mathbf{u} \label{eq:clf_derivative}
    \end{align}
    As in our case $a(y) = 0$ a stabilizing controller is simply given by:
    \begin{align}
        \mathbf{u^{c}} = \mathbf{-b} \label{app-eq:control_easy}
    \end{align} 
    which then results in the following derivative for $V$:
    \begin{align}
        \dot{V} = -\mathbf{b}^2 \leq 0
    \end{align}
    and thus shows that $V$ is a CLF up to the condition that $\dot V$ must not become zero away from the origin.

    We present the following CLF $V$ for our system:
    \begin{align}
        V(S, \Sbar, \nu) &= \sum_h |S_h - S^d_h|^2 + (\nu_h - \nu_h^d)^2 \geq 0 \\
        &= \sum_h \Delta S_h \cdot \Delta \Sbar_h + \Delta \nu_h \cdot \Delta \nu_h. \\
    \end{align}

    Using equation \eqref{eq:power_coupling} and \eqref{eq:nu_dynamics} to get the derivatives of $S_h$ and $\nu_h$ respectively we get the following expression for $\dot V$:
    \begin{align}        
        \dot{V}(S, \Sbar, \nu) &= \sum_h \dot{S}_h \cdot \Delta \Sbar_h + \Delta S_h \cdot \dot{\Sbar}_h + 2 \dot{\nu}_h \cdot \Delta \nu_h\\
        &= \sum_h S_h \eta_h \Delta \Sbar_h + \Delta \Sbar_h \sum_m K_{hm} \etabar_m + \Delta S_h \etabar_h \Sbar_h + \Delta S_h \sum_m \Kbar_{hm} \eta_m + 2 \Delta \nu_h \nu_h (\eta_h + \etabar_h)
    \end{align}
    We rearrange the equation to isolate the control inputs $\eta_h$ and $\etabar_h$:
    \begin{align}        
        \dot{V} &= \sum_h \eta_h (S_h \Delta \Sbar_h + 2 \Delta \nu_h \nu_h) + \etabar_h (\Delta S_h \etabar_h \Sbar_h + 2 \Delta \nu_h \nu_h) + \sum_h  \sum_m K_{hm} \etabar_m \Delta \Sbar_h +  \Delta S_h \Kbar_{hm} \eta_m. 
    \end{align}
    After switching the indices $m$ and $h$ in the second sum we obtain $\dot{V}$ in the form of equation \eqref{eq:clf_derivative} which allows us to directly define the control:
    \begin{align}        
        \dot{V} &= \sum_h \eta_h (S_h \Delta \Sbar_h + 2 \Delta \nu_h \nu_h + \sum_m \Delta S_m \Kbar_{mh}) + \etabar_h (\Delta S_h \Sbar_h + 2 \Delta \nu_h \nu_h + \sum_m K_{mh} \Delta \Sbar_m) \\
        &= \sum_h \eta_h \overline{b}_h + \etabar_h b_h. \\
        b_h &= \Delta S_h \Sbar_h + 2 \Delta \nu_h \nu_h + \sum_m K_{mh} \Delta \Sbar_m
    \end{align}
    Using the control law \eqref{app-eq:control_easy} we define the control for $\eta^{c}$ as:
    \begin{align}
        \eta_h = \eta_{h}^{c} := - b = \Delta S_h \Sbar_h + 2 \Delta \nu_h \nu_h + \sum_m K_{mh} \Delta \Sbar_m
    \end{align}
    then $\dot{V}$ is given by:
    \begin{align}
        \dot V &= - 2 \sum_h \etabar_{h}^{c} \cdot \eta^{c}_h \leq 0
    \end{align}
    which shows that $V$ is a CLF and $\eta^{c}$ results in global asymptotic stability. For both, the power flow and the coupling dynamics, the state space is larger than the space of physically realizable flows. There is no requirement that a voltage $v_h$ exists that can realize the $K_{hm}$ given. However, if we start on the physical manifold, then by construction the dynamics driven by the complex frequency will stay on that manifold. The calculation showing this can be found in section \ref{sec:voltage_existence}.

    \subsection{Example Trajectories}
    \label{app:lcf_examples}
\begin{figure}[H]
            \centering
            \begin{subfigure}[b]{0.32\textwidth}
                \centering
                \includegraphics[width=\textwidth]{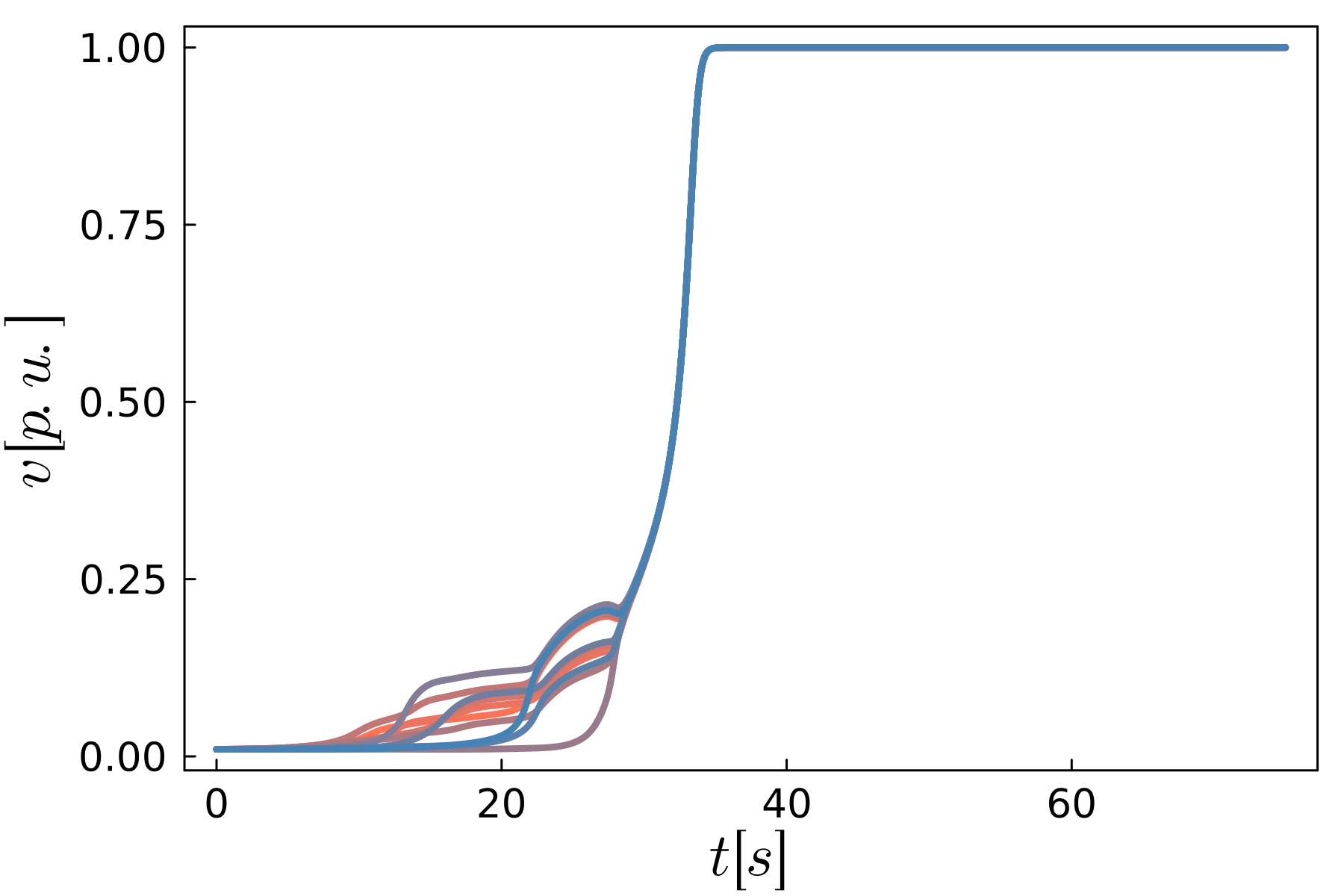}
            \end{subfigure}
            \hfill
            \begin{subfigure}[b]{0.32\textwidth}
                \centering
                \includegraphics[width=\textwidth]{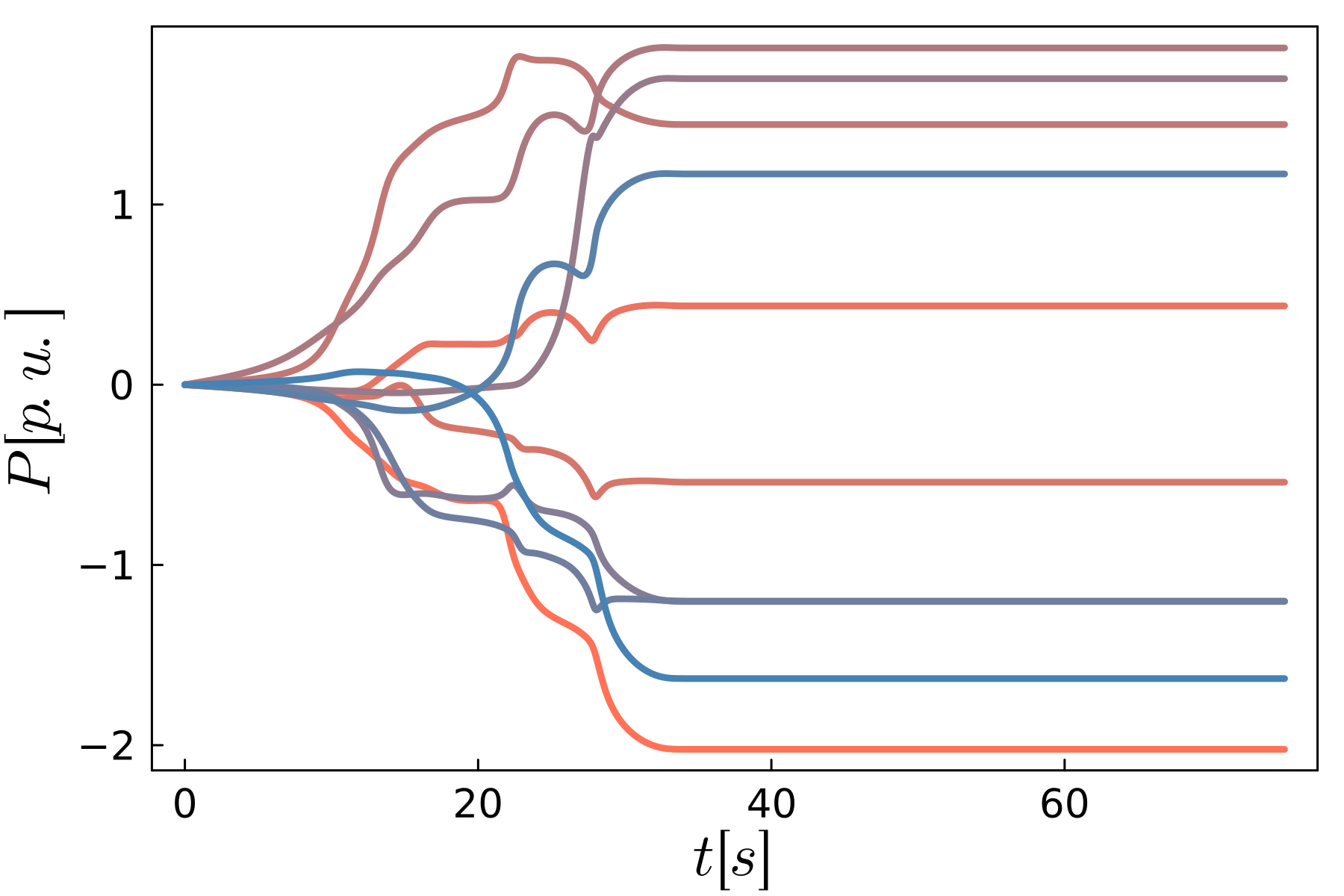}
            \end{subfigure}
            \hfill
            \begin{subfigure}[b]{0.32\textwidth}
                \centering
                \includegraphics[width=\textwidth]{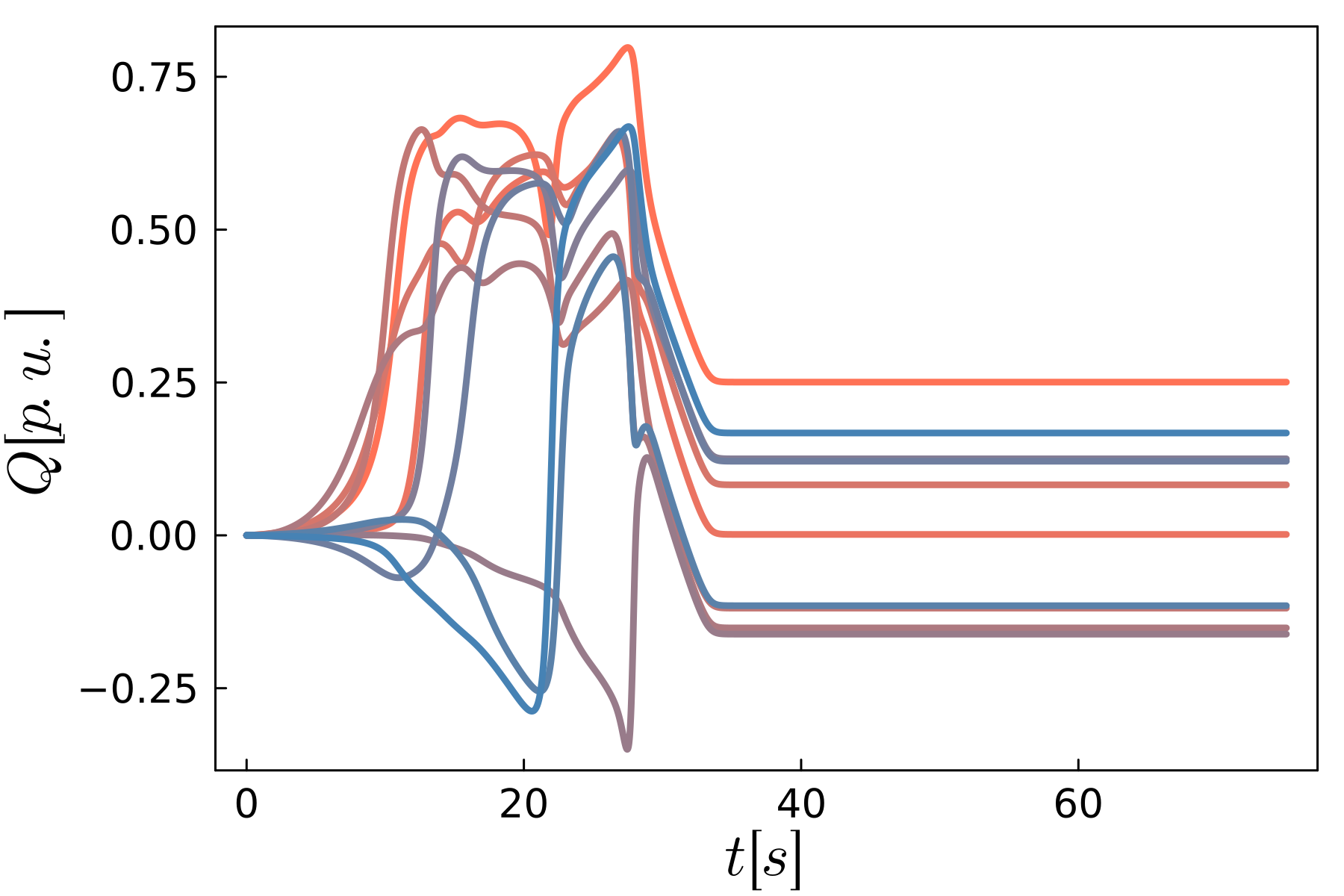}
            \end{subfigure}
            \caption{Voltage magnitude, active and reactive power transients during the black start of the 10 node test system.}
            \label{fig:blackstart}
        \end{figure}

    To study a more severe control task we also consider a black start of the 10-bus system, which means that we decrease all voltages to $0.01 [p.u.]$. We used this voltage magnitude as a voltage of $0.0 [p.u.]$ is a fixed point of the dynamical system as discussed in section \ref{sec:bilinear_coupling}. From figure \ref{fig:blackstart} we can see that our control can perform a black start for the test system in 35 seconds.

\begin{figure}[H]
            \centering
            \includegraphics[width=0.5\textwidth]{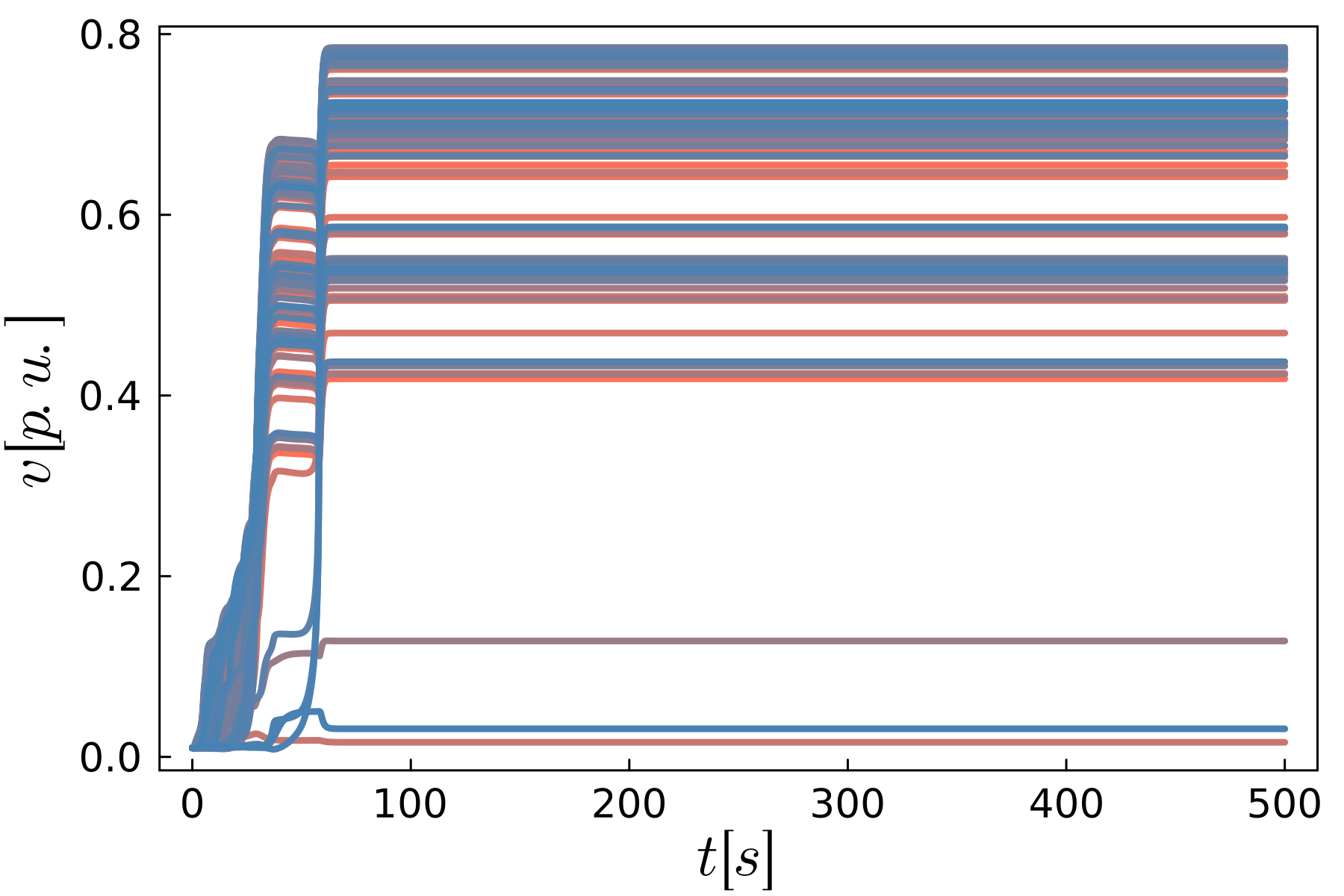}
            \caption{Attempted black start of a system with 100 nodes.}
            \label{fig:blackstart_100n}
        \end{figure}

        However, the black start capabilities are not given for arbitrary systems. Figure \ref{fig:blackstart_100n} shows an attempted black start of a 100-node system generated by the algorithm given in \cite{buettner_synthetic_power_grids_2023}. It can be seen that the system voltages are not returning to their nominal values. The test system returns to another fixed point that is not the operating point of the grid.
   \subsection{Existence of the voltage}
   \label{sec:voltage_existence}
    For both, the power flow and the coupling dynamics, the state space is larger than the space of physically realizable flows. There is no requirement that a $v_h$ exists that can realize the $K_{hm}$ given. However, if we start on the physical manifold, then by construction the dynamics driven by the complex frequency will stay on that manifold.  The calculation showing this can be found in the appendix.
    To understand where the choice of $\eta^{c}$ we consider the local minima of $V$ on the physical manifold.
    \begin{align}
        \min_{S, \Sbar, \nu, \lambda^S, \lambdabar^S, \lambda^\nu, v, \vbar} \left( V(S, \Sbar, \nu) - \lambdabar^S \cdot (S - [v] \Ybar \vbar) - \lambda^S \cdot (S - [\vbar] Y v) - \lambda^\nu \cdot (\nu - [v] \vbar) \right)
    \end{align}
    The variation are then given by:
    \begin{align}
        \partial_S :& \Delta \Sbar - \lambdabar^S = 0 \\
        \partial_\Sbar :& \Delta S - \lambda^S = 0 \\
        \partial_\nu :& 2 \Delta \nu - \lambda^\nu = 0 \\
        \partial_v :& [\lambdabar^S] \Ybar \vbar - Y^T [\vbar] \lambda^S + [\lambda^\nu] \vbar = 0 \\
        \partial_\vbar :& \text{ c.c. }
    \end{align}

    Multiplying the $\partial_v$ variation with $[v]$, and noting that we have $[v] \Ybar \vbar = S$ and $[v] Y^T [\vbar] = \Kbar^T$ we see that:

    \begin{align}
        [v] \partial_v : 0 &= [\lambdabar^S] [S] - \Kbar^T \lambda^S + [\lambda^\nu] [v] \vbar \\
        & = [\Delta \Sbar] S - \Kbar^T \Delta S + 2 [\Delta \nu] \nu \\
        & = \etabar^{c}
    \end{align}
    Thus with this quasi-local control (the control depends on the power imbalance at the neighbors and the state of the coupling on the edges), $V$ is a Lyapunov function. If the set points satisfy the power flow equations, then the system at the stable power flow has $V=0$ and is Lyapunov stable. The control law $\eta^{c}$ always drives the system to the local minima of $V$ on the physical manifold, and these minima are fixed points as the local minima satisfy $\eta^{c} = 0$.

\end{document}